\documentclass{iopart}
\usepackage{euscript,iopams,amssymb,amsfonts,graphicx,bm}
\usepackage{pgfplots,setstack}

\usepackage{float}
\usepackage{epsfig}
\usepackage{esint}

\usepackage{bm,braket}

\eqnobysec

\newtheorem{theorem}{Theorem}[section]

\renewcommand{\v}{{\mathbf v}}

\newcommand{\p}{{\mathbf p}}

\newcommand{\R}{{\mathbb R}}

\renewcommand{\L}{{\mathbb L}}

\renewcommand{\e}{\mathrm{e}}

\newcommand{\x}{\mathbf{x}}

\renewcommand{\P}{\mathbb{P}}

\renewcommand{\theequation}{\arabic{section}.\arabic{equation}}

\newcommand{\E}{{\mathbb E}}
\newcommand{\Z}{\mathbf Z}

\begin{document}
\title[Stochastic hybrid path integrals]{Construction of stochastic hybrid path integrals using ``quantum-mechanical'' operators}
\author{Paul C. Bressloff}
\address{Department of Mathematics, University of Utah, Salt Lake City, UT, USA} \ead{bressloff@math.utah.edu}

\begin{abstract}
Stochastic hybrid systems involve the coupling between discrete and continuous stochastic processes. They are finding increasing applications in cell biology, ranging from modeling promoter noise in gene networks to analyzing the effects of stochastically-gated ion channels on voltage fluctuations in single neurons and neural networks. We have previously derived a path integral representation of solutions to the associated differential Chapman-Kolmogorov equation, based on integral representations of the Dirac delta function, and used this to determine ``least action'' paths in the noise-induced escape from a metastable state. In this paper we present an alternative derivation of the path integral, based on the use of bra-kets and ``quantum-mechanical'' operators. We show how the operator method provides a more efficient and flexible framework for constructing hybrid path integrals, which eliminates certain ad hoc steps from the previous derivation and provides more context with regards the general theory of stochastic path integrals. We also highlight the important role of principal eigenvalues, spectral gaps and the Perron-Frobenius theorem. We then use perturbation methods to develop various approximation schemes for hybrid path integrals and the associated moment generating functionals. First, we consider Gaussian approximations and loop expansions in the weak noise limit, analogous to the semi-classical limit for quantum path integrals. Second, we identify the analog of a weak-coupling limit by treating the stochastic hybrid system as the nonlinear perturbation of an Ornstein-Uhlenbeck process. This leads to an expansion of the moments in terms of products of free propagators.
\end{abstract}

\maketitle


\section{Introduction}

Path integral methods \cite{Feynman48,Kleinert09,Atland10,Kamenev14} have become an increasingly useful tool for studying continuous time discrete stochastic processes that evolve according to a master equation. Applications include chemical kinetics \cite{Elgart04,Tauber05,Elgart06}, gene networks \cite{Aurell02a,Sasai03,Roma05,Li14}, population dynamics \cite{Assaf06,Kamenev08,Assaf09}, and neural networks \cite{Buice07,Bressloff09,Buice10}. For an extensive review and list of references see \cite{Weber17}. A path integral representation of the solution to a master equation was originally derived by Doi and Peliti \cite{Doi76,Doi76a,Peliti85} using an operator method that borrows from the bra-ket formalism of quantum mechanics. This followed on the heels of a corresponding path integral formulation of stochastic differential equations (SDEs), which was derived using an integral representation of the Dirac delta function \cite{Martin73,Dom76,Janssen76}. The latter approach, which avoids the use of operators, has recently been extended to master equations by considering differential equations for the corresponding generator or marginalized distribution of the Markov process \cite{Weber17}. On the other hand, Vastola and Holmes \cite{Holmes20} have shown how the path integral for SDEs can also be derived using a bra-ket description, suggesting that ``quantum mechanical'' operators provide a general tool for constructing path integrals of stochastic processes. 

In this paper we use the bra-ket formalism and operator methods to derive a path integral for stochastic hybrid systems, which combine discrete and continuous stochastic processes. There are a growing number of examples in biological physics that are modeled in terms of stochastic hybrid systems \cite{Bressloff14a,Bressloff17a}. These include conductance-based neuron models \cite{Fox94,Chow96,Keener11,Goldwyn11,Buckwar11,NBK13,Bressloff14b,Newby14}, where a population of membrane-bound ion channels stochastically open and close with transition rates that depend on the membrane voltage. The dynamics of the latter itself depends on the current state of the ion channels. Hence, one can identify the number of open ion channels at time $t$ as a discrete random variable $N(t)$ and the voltage as a continuous random variable $X(t)$. Another major application is the study of promoter noise in gene networks, where $N(t)$ represents the activity state of a gene (due to the binding/unbinding of transcription factors) and $X(t)$ is the number of mRNA or synthesized proteins \cite{Kepler01,Bose04,Newby12,Newby15,Hufton16}. A final example is a stochastic hybrid neural network that is modeled in terms of a set of synaptically coupled neuronal populations \cite{Bressloff13a,Bressloff15,Yang19}. The state of each local population is described in terms of two stochastic variables, a continuous synaptic variable and a discrete activity variable.

A special type of stochastic hybrid system is a so-called piecewise deterministic Markov process (PDMP), in which the continuous random variables evolve deterministically between jumps in the discrete random variables \cite{Davis84}. These have been studied extensively within the context of large deviation theory \cite{Kifer09,fagg09,Faggionato10,Bressloff17}. We have previously derived a path integral formulation of PDMPs using integral representations of Dirac delta functions \cite{Bressloff13a,Bressloff15}, analogous to the analysis of SDEs in Refs. \cite{Martin73,Dom76,Janssen76}. However, the derivation was rather involved, and some of the steps appeared a little ad hoc. In particular, we introduced an eigenvalue equation that included an arbitrary parameter, which was subsequently fixed to be the ``momentum'' variable within the path integral. The leading or principal eigenvalue was then identified as the effective Hamiltonian of the path integral action. In this paper we show how the same path integral representation can be derived much more cleanly using operator methods. The emergence of the principal eigenvalue occurs naturally without the need to introduce any arbitrary parameters. We also generalize the path integral construction to include intrinsic noise within the piecewise continuous dynamics.

The structure of the paper is as follows. In section 2, we review the Doi-Peliti creation and annihilation operator formalism for a birth-death process, and discuss some alternative formulations that are more useful for stochastic hybrid systems. We then briefly describe the operator method for SDEs that was recently introduced in \cite{Holmes20}. Finally, we combine the two cases to develop a corresponding operator formalism for stochastic hybrid systems. The latter is then used to construct the hybrid path integral in section 3. In section 4, we highlight the important role of principal eigenvalues, spectral gaps and the Perron-Frobenius theorem. We discuss a few examples for which the principal eigenvalue can be calculated explicitly. We also indicate how to estimate the principal eigenvalue using a Ritz variational method, analogous to calculating the energy ground state of a quantum system. Finally, in section 5 we use perturbation methods to develop various approximation schemes for hybrid path integrals and the associated moment generating functionals. First, we consider Gaussian approximations and loop expansions in the weak noise limit, analogous to the semi-classical limit for quantum path integrals. Second, we identify the analog of a weak-coupling limit by treating the stochastic hybrid system as the nonlinear perturbation of an Ornstein-Uhlenbeck process. This leads to an expansion of the moments in terms of products of free propagators.

\section{Bra-ket formulation of stochastic processes}

 In this section we show how the bra-ket formulations of master equations \cite{Doi76,Doi76a,Peliti85} and SDEs \cite{Holmes20} can be combined to provide a corresponding operator formulation of a stochastic hybrid system. 

\subsection{Birth-death master equation}

In order to illustrate the bra-ket representation of a chemical master equation, we focus on the relatively simple example of a birth-death process for a discrete random variable $N(t)\in {\mathbb Z}^+$. (The construction also extends to more general master equations with multiple rather than single step reactions, although the analysis tends to become more cumbersome.) Setting $P(n,t)=\P[N(t)=n]$, the birth-death master equation takes the general form
\begin{eqnarray}
\label{bdbd}
\fl \frac{dP(n,t)}{dt}&=\sum_{m\geq 0}Q_{nm}P(m,t)\equiv \omega_+(n-1)P(n-1,t)
+  \omega_-(n+1)P(n+1,t)\nonumber \\
\fl &\quad -[\omega_+(n)+ \omega_-(n)]P(n,t), 
\end{eqnarray}
with boundary condition $P(-1,t)=0$ and birth/death rates $\omega_{+}(n),\omega_-(n)$. The initial condition is $P_n(0)=\sigma_n$ with $\sum_n\sigma_n=1$.

\subsubsection{Doi-Peliti operator formalism} The starting point of the operator formalism developed by Doi and Peliti is to introduce an abstract ``bosonic'' vector space (also known as a Fock space) with elements $|n\rangle$ representing the discrete states, together with a pair of creation--annihilation linear operators that satisfy the commutation relation
\numparts
\begin{equation}
\label{com1}
[a,a^{\dagger}]\equiv aa^{\dagger}-a^{\dagger}a=1.
\end{equation}
These operators generate the full vector space by acting on the ``vacuum'' state $|0\rangle$, with $a|0\rangle=0$. The state $|n\rangle$ is then generated according to
\begin{equation}
|n \rangle ={a^{\dagger}}^{n}|0\rangle.
\end{equation}
Inner products in this state space are defined by $\langle 0|0 \rangle =1$ and the commutation relation. It follows that the dual of the vector ${a^{\dagger}}|0\rangle$ is $\langle 0|a$ and $\langle n|m\rangle=\delta_{n,m}n!$. Other standard operator equations are
\begin{equation}
a|n\rangle =n|n-1\rangle,\ a^{\dagger}|n\rangle = |n+1\rangle,\ a^{\dagger}a|n\rangle  =n|n\rangle .
\end{equation}
\endnumparts
In addition, the basis vectors $|n\rangle$ satisfy the completeness relation (or resolution of the identity)
\begin{equation}
\label{bdcom1}
\sum_{n\geq 0}\frac{1}{n!}|n\rangle \langle n|=1.
\end{equation}
That is, for an arbitrary vector $|c\rangle =\sum_{m\geq 0}c_m|m\rangle$,
\begin{eqnarray*}
\sum_{n\geq 0}\frac{1}{n!}|n\rangle \langle n|c\rangle&=\sum_{n,m\geq 0}\frac{1}{n!}|n\rangle c_m\langle n|m\rangle=\sum_{n,m\geq 0}\frac{1}{n!}|n\rangle c_m m!\delta_{n,m}\\
&=\sum_{m\geq 0}c_m|m\rangle=|c\rangle.
\end{eqnarray*}

The next step is to construct an operator representation of the master equation (\ref{bdbd}). Given the probability distribution $P(n,t)$, we define the state vector by
\begin{equation}
|\psi(t)\rangle = \sum_{n\geq 0}P(n,t) {a^{\dagger}}^{n}|0\rangle=\sum_{n\geq 0}P(n,t)|n\rangle .
\end{equation}
Introducing the projection state
\begin{equation}
| \emptyset\rangle= \exp\left (a^{\dagger}\right )|0\rangle=\sum_{n=0}^{\infty} \frac{1}{n!}\ket{n},
\end{equation}
with $a|\emptyset \rangle = |\emptyset\rangle$ and $\langle \emptyset|m\rangle =1$,
expectation values can be expressed in terms of inner products. For example,
\begin{equation}
\langle \emptyset|a^{\dagger}a |\psi(t)\rangle = \sum_{n=0}^{\infty}nP(n,t)=\langle N(t)\rangle.
\end{equation}
Differentiating the state vector $|\psi(t)\rangle$ with respect to $t$ and using the master equation (\ref{bdbd}) one obtains the operator equation
\numparts
\begin{equation}
 \label{op1}
\frac{d}{dt} |\psi(t)\rangle=\hat{H}_{\rm bd} |\psi(t)\rangle ,
\end{equation}
with
\begin{equation}
\label{Hcal}
\hat{H}_{\rm bd}=(a- a^{\dagger}a)\overline{\omega}_-(a^{\dagger}a)   +(a^{\dagger}-1)\omega_+(a^{\dagger}a),
 \end{equation}
 \endnumparts
 where $n\overline{\omega}_-(n)=\omega_-(n)$.
Formally speaking, the solution to the operator version of the master equation (\ref{op1}) is
 \begin{equation}
 \label{eq2:phi}
|\psi(t)\rangle =\e^{\hat{H}_{\rm bd}t}|\psi(0)\rangle ,
\end{equation}
 and the expectation value of some physical quantity such as the number $N(t)$ takes the form 
 \begin{equation}
\langle N(t)\rangle =\langle \emptyset| a^{\dagger}a \e^{\hat{H}_{\rm bd}t}|\psi(0)\rangle .
\end{equation}

The final ingredient of the bra-ket formalism is the choice of basis vectors. For example, the construction of the Doi-Peliti path integral works with the coherent-state representation
 \begin{equation}
 |\varphi\rangle = \exp\left (-\frac{1}{2}|\varphi|^2 \right )\exp\left (\varphi a^{\dagger}\right )|0\rangle,
 \end{equation}
where $\varphi$ is the complex-valued eigenvalue of the annihilation operator $a$, with complex conjugate ${\varphi}^*$. Coherent states satisfy the completeness relation
\begin{equation}
\label{bdcom2}
\int\frac{d\varphi d{\varphi}^* }{\pi}|\varphi\rangle \langle \varphi|=1.
\label{comp}
\end{equation}
In order to determine the action of $\hat{H}_{\rm bd}$ on $|\varphi\rangle$ it is first necessary to normal-order $\hat{H}_{\rm bd}$ by moving all creation operators to the left of all annihilation operators using the commutation relation. For example,
if $w_+(n)=n^2$ then 
\begin{eqnarray*}
\fl \omega_+(a^{\dagger}a) =a^{\dagger}aa^{\dagger}a=a^{\dagger}[a,a^{\dagger}]a+a^{\dagger}a^{\dagger}aa=a^{\dagger}a+a^{\dagger}a^{\dagger}aa
=[{\omega}]_+(a,a^{\dagger}),
\end{eqnarray*}
where $[\omega]$ denotes the transition rate after normal-ordering.
It follows that
\begin{equation}
\langle \varphi| \hat{H}_{\rm bd}|\varphi\rangle =
(\varphi- \varphi^*\varphi)[\overline{\omega}]_-(\varphi,\varphi^*)   +(\varphi^*-1)[{\omega}]_+(\varphi,\varphi^*).
\end{equation}

Another basis can be constructed when the birth-death operator $\hat{H}_{\rm bd}$ has a discrete spectrum whose corresponding eigenfunctions form a complete orthogonal set for the underlying Fock space. This is particularly useful when the discrete state-space is finite, rather than unbounded. Let $\lambda_{\mu}$ denote the $\mu$-th eigenvalue with associated eigenvector $|r_{\mu}\rangle$:
\begin{equation}
\label{eig:bd}
\hat{H}_{\rm bd}|r_{\mu}\rangle = \lambda_{\mu}|r_{\mu}\rangle,\quad |r_{\mu}\rangle =\sum_{n\geq 0}r_{\mu}(n)|n\rangle .
\end{equation}
It is important to note that the operator $\hat{H}_{\rm bd}$ is non-Hermitian, which means its right and left (or dual) eigenvectors are not simply adjoints of each other. More specifically,  
\begin{equation}
\langle \bar{r}_{\mu}|\hat{H}_{\rm bd} =\lambda_{\mu}\langle \bar{r}_{\nu}|,\quad \langle \bar{r}_{\mu} | =\sum_{n\geq 0}\frac{\bar{r}_{\mu}(n)}{n!}\langle n|,
\end{equation}
such that
\begin{equation}
\langle \bar{r}_{\mu} |r_{\nu}\rangle = \sum_{n\geq 0} \bar{r}_{\mu}(n)r_{\nu}(n)=\delta_{\mu,\nu}.
\end{equation}
One also has the completeness relation
\begin{equation}
\label{bd:comp3}
\sum_{\mu \geq 0}|r_{\mu}\rangle\langle\bar{r}_{\mu}|=1.
\end{equation}
This follows from the corresponding completeness relation
\begin{equation}
 \sum_{\mu\geq 0} \bar{r}_{\mu}(m)r_{\mu}(n)=\delta_{m,n},
 \end{equation}
since
\begin{eqnarray*}
\fl &\sum_{\mu \geq 0}|r\rangle\langle \bar{r}|=\sum_{\mu \geq 0}\sum_{m,n\geq0}
\frac{1}{m!}\bar{r}_{\mu}(m)r_{\mu}(n)|n\rangle \langle m|=\sum_{m,n\geq0}\frac{1}{m!}|n\rangle \langle m|\sum_{\mu \geq 0}\bar{r}_{\mu}(m)r_{\mu}(n)\\
\fl&=\sum_{\mu \geq 0}\sum_{m,n\geq0}\frac{1}{m!}|n\rangle \langle m|\delta_{m,n} =\sum_{n\geq0}\frac{1}{n!} |n\rangle \langle n|=1.
\end{eqnarray*}
(We are assuming that if the number of discrete states is infinite, then it is possible to change the summation order.) 
Moreover, in the original basis,
\begin{eqnarray*}
\fl \hat{H}_{\rm bd}|\mu\rangle &=\sum_{n\geq 0}r_{\mu}(n)\hat{H}_{\rm bd}|n\rangle=\sum_{n\geq 0}\bigg [\omega_+(n-1)r_\mu(n-1) +  \omega_-(n+1)r_\mu(n+1)\nonumber \\
\fl &\hspace{4cm} -[\omega_+(n)+\omega_-(n)]r_{\mu}(n)\bigg]|n\rangle .
\end{eqnarray*}
Taking the inner product of the operator eigenvalue equation with respect to $\langle m|$ then yields the matrix eigenvalue equation
\begin{eqnarray}
\label{Qr}
\sum_{m\geq 0}Q_{nm}r_{\mu}(m)=\lambda_{\mu} r_{\mu}(n). \label{eq2:eig0}
\end{eqnarray}
That is, the vector with components $r_{\mu}(n)$ is a (right) eigenvector of the birth-death matrix generator.

\subsubsection{Alternative operator pair} In the case of single step reaction schemes such as birth-death processes, a simpler operator formalism can be developed as follows \cite{Atland10}. First note that the birth-death master equation
can be rewritten in the form
\begin{eqnarray}
\label{bd0}
\frac{dP(y,t)}{dt}&=\left [(\E_1-1)\omega_-(\overline{N}y)+(\E_{-1}-1)\omega_+(\overline{N}y)\right ]P(y,t),
\end{eqnarray}
where $\E_{\pm 1}f(y)=f(y\pm 1/\overline{N})$ and $\overline{N}$ is a system size. As in the system-size expansion of master equations, we treat $y=n/\overline{N}$ as a continuous variable $y\in \R^+$ and consider a Hilbert space spanned by the vectors $|y\rangle$ with inner product and completeness relation
\begin{equation}
\label{zcom0}
\langle y'|y\rangle =\delta(y-y'),\quad \int_0^{\infty}dy|y\rangle \langle y|=1.
\end{equation}
(If $\overline{N}=1$ the above construction is still valid, since the operators $\E_{\pm 1}$ shift $y$ by $\pm 1$. Hence, if $y$ is initially an integer, then it remains an integer.)
Introduce the conjugate pair of operators $\hat{y},\hat{q} $ such that
\begin{equation}
\hat{y}|y\rangle =\overline{N}y|y\rangle,\quad \hat{q} =-\frac{1}{\overline{N}}\overset{\leftarrow}{\frac{\partial}{\partial y}}|y\rangle ,\quad [\hat{y},\hat{q}]=1.
\end{equation}
The arrow on the differential operator indicates that it operates to the left. This is equivalent to defining the action of $\hat{q}$ on general state vectors $|\phi\rangle =\int_0^{\infty}dy \phi(y)|y\rangle$:
\begin{eqnarray*}
\fl \hat{q}|\phi\rangle=\hat{q}\int_0^{\infty}dy \phi(y)|y\rangle=-\frac{1}{\overline{N}}\int_0^{\infty}dy \phi(y)\left [\overset{\leftarrow}{\frac{\partial}{\partial y}} \right ]|y\rangle=-\frac{1}{\overline{N}}\int_0^{\infty}dy \phi'(y)]|y\rangle .
\end{eqnarray*}
That is, $\langle y|\hat{q}|\phi\rangle = -\phi'(y)/\overline{N}$.

Introducing the state vector $|\psi(t)\rangle =\int_0^{\infty}dy P(y,t)|y\rangle$, the master equation can be rewritten in the operator form
\begin{equation}
\label{H'}
\fl  \frac{d}{dt}|\psi(t)\rangle = \hat{H}_{\rm bd}'|\psi(t)\rangle,\quad \hat{H}_{\rm bd}' =(\e^{-\hat{q} }-1) \omega_-(\hat{y})+(\e^{\hat{q}}-1)\omega_+(\hat{y}),
\end{equation}
which again has the formal solution $|\psi(t)\rangle =\e^{\hat{H}_{\rm bd}'t}|\psi(0)\rangle$. Using the fact that the ``momentum'' vector
\begin{equation}
|q\rangle=\int_0^{\infty}dy\e^{-qy}|y\rangle
\end{equation}
is an eigenvector of $\hat{q} $ with eigenvalue $q$, we have
\begin{equation}
\label{eq8:H'}
\langle q|\hat{H}_{\rm bd}' |y\rangle =H_{\rm bd}'(y,q)\equiv  (\e^{-q}-1)\omega_-(\overline{N}y)+(\e^{q}-1)\omega_+(\overline{N}y).
\end{equation}
Analogous to taking an inverse Laplace transform using a Bromwich integral, the corresponding completeness relation for $|q\rangle$ is \cite{Holmes20}
\begin{equation}
\label{zcom}
\int_{-i\infty}^{i\infty}\frac{dq}{2\pi} |q\rangle \langle q|=1.
\end{equation}
Finally, note that $\hat{H}_{\rm bd}' $ is equivalent to $\hat{H}_{\rm bd} $ of equation (\ref{Hcal}) under the mapping
\begin{equation}
\label{eq8:CH}
a=\e^{-\hat{q} }\hat{y},\quad a^{\dagger}=\e^{\hat{q}},
\end{equation}
which is a form of Cole-Hopf transformation \cite{Atland10}.

\subsection{Fokker-Planck equation} Consider a  continuous stochastic process $X(t)\in \R$ evolving according to the Ito SDE
\begin{equation}
dX(t)=A(X)dt+\sqrt{2D(X)}dW(t),
\end{equation}
where $W(t)$ is a Wiener process with
\begin{equation}
\langle W(t)\rangle =0,\quad \langle W(t)W(s)\rangle =\min\{t,s\}.
\end{equation}
The corresponding FP equation for the probability density $P(x,t)$ is
\begin{equation}
\frac{\partial P(x,t)}{\partial t}=-\frac{\partial A(x)P(x,t)}{\partial x}+\frac{\partial^2 D(x)P(x,t)}{\partial x^2}.
\end{equation}
\numparts
Following Ref. \cite{Holmes20}, we introduce a Hilbert space spanned by the vectors $|x\rangle$, together with a conjugate pair of position-momentum operators $\hat{x}$ and $\hat{p}$ such that
\begin{equation}
\label{com2}
[\hat{x},\hat{p}]=i.
\end{equation}
Their action on the given Hilbert space is taken to be
\begin{equation}
\hat{x}|x\rangle = x|x\rangle,\quad \hat{p}|x\rangle=-i\overset{\leftarrow}{\frac{d}{d x}}|x\rangle.
\end{equation}
Again the arrow on the differential operator indicates that it operates to the left. Alternatively, given a state vector $|\phi\rangle =\int_{-\infty}^{\infty}dx\phi(x) |x\rangle$, we have $\langle x|\hat{p}|\phi\rangle =-i\phi'(x)$.
 \endnumparts
These operators satisfy the commutation relation since
\begin{eqnarray*}
[\hat{x},\hat{p}]|x\rangle &=\hat{x}\hat{p}|x\rangle
- \hat{p}\hat{x}|x\rangle=\hat{x}\left (-i\overset{\leftarrow}{\frac{d}{d x}}\right)|x\rangle
-\hat{p}x|x\rangle\\
&=\left (-i\overset{\leftarrow}{\frac{d}{d x}}\right)x|x\rangle
-x\left (-i\overset{\leftarrow}{\frac{d}{d x}}\right )|x\rangle=i|x\rangle.
\end{eqnarray*}
The inner product and completion relations on the Hilbert space are
\begin{equation}
\label{fpcom1}
\langle x'|x\rangle =\delta(x-x'),\quad  \int_{-\infty}^{\infty} dx\, |x\rangle \langle x|=1.
\end{equation}
 That is, for an arbitrary vector $|c\rangle =  \int_{-\infty}^{\infty} dx\,c(x)|x\rangle$,
 \begin{eqnarray*}
\fl \int_{-\infty}^{\infty} dx\, |x\rangle \langle x|c\rangle&=\int_{-\infty}^{\infty} dx\,\int_{-\infty}^{\infty} dy\,   |x\rangle c(y) \langle x|y\rangle\\
\fl &=\int_{-\infty}^{\infty} dx\,\int_{-\infty}^{\infty} dy\,   |x\rangle c(y)\delta(x-y)=\,\int_{-\infty}^{\infty} dy\, c(y)|y\rangle = |c\rangle.
 \end{eqnarray*}

Given the probability density $P(x,t)$ we define the state vector 
\begin{equation}
|\psi(t)\rangle = \int_{-\infty}^{\infty} dx\, P(x,t)|x\rangle .
\end{equation}
Differentiating both sides with respect to time $t$ and using the FP equation gives
\begin{eqnarray*}
\frac{d}{dt}|\psi(t)\rangle&= \int_{-\infty}^{\infty} dx\,\left [-\frac{\partial A(x)P(x,t)}{\partial x}+\frac{\partial^2 D(x)P(x,t)}{\partial x^2}\right ]|x\rangle\\
&= \int_{-\infty}^{\infty} dx\,\left [-A(x)P(x,t)\overset{\leftarrow}{\frac{\partial }{\partial x}}+ D(x)P(x,t)\overset{\leftarrow}{\frac{\partial^2}{\partial x^2}}\right ]|x\rangle\\
&=\left [-i\hat{p}A(\hat{x})-\hat{p}^2 D(\hat{x})\right ]\int_{-\infty}^{\infty} dx\, P(x,t)|x\rangle.
\end{eqnarray*}
Hence, we can write the FP equation in the operator form
\numparts
\begin{equation}
\frac{d}{dt}|\psi(t)\rangle=\hat{H}_{\rm fp} |\psi(t)\rangle,
\end{equation}
with
\begin{equation}
\label{fp:H}
\hat{H}_{\rm fp}=-i\hat{p}A(\hat{x})-\hat{p}^2 D(\hat{x}).
\end{equation}
\endnumparts
The formal solution of the FP equation is
\begin{equation}
|\psi(t)\rangle=\e^{\hat{H}_{\rm fp}t}|\psi(0)\rangle,
\end{equation}
and expectations are given by
\begin{equation}
\fl \langle X(t)\rangle = \int_{\infty}^{\infty}dx\, xP(x,t)=\int_{-\infty}^{\infty} dx\, \langle x|\hat{x}|\psi(t)\rangle=\int_{-\infty}^{\infty} dx\, \langle x|\hat{x}\e^{\hat{H}_{\rm fp}t}|\psi(0)\rangle.
\end{equation}

As in the case of the Doi-Peliti construction, one can consider different choices of basis vectors. The most natural alternative to $|x\rangle$ is the momentum representation (analogous to taking Fourier transforms),
\begin{equation}
|p\rangle =\int_{-\infty}^{\infty}dx\,  \e^{ipx}|x\rangle.
\end{equation}
It immediately follows that $|p\rangle$ is an eigenvector of the momentum operator $\hat{p}$, since
\begin{equation}
\hat{p}|p\rangle=\int_{-\infty}^{\infty}dx\,  \e^{ipx}\left (-i\overset{\leftarrow}{\frac{d}{d x}}\right )|x\rangle=\int_{-\infty}^{\infty}dx\, p \e^{ipx}|x\rangle =p|p\rangle.
\end{equation}
Using the inverse Fourier transform, we also have
\begin{equation}
|x\rangle =\int_{-\infty}^{\infty}\frac{dp}{2\pi}\,  \e^{-ipx}|p\rangle,
\end{equation}
and the completeness relation
\begin{equation}
\label{fpcom2}
\int_{-\infty}^{\infty}\frac{dp}{2\pi}|p\rangle \langle p|=1.
\end{equation}

\subsection{Stochastic hybrid system} 

The bra-ket formulations of a master equation and an FP equation can now be combined to develop a corresponding treatment of the Chapman-Kolmogorov (CK) equation for a stochastic hybrid system. Let the state of the system at time $t$ consist of the pair $(X(t),N(t))$, where $X(t)\in \R$ and $N(t)\in {\mathbb Z}^+$. Suppose that the discrete process evolves according to a birth-death process of the form (\ref{bdbd}), except now the transition rates may depend on the continuous state variable $X(t)$. That is,
$\omega_{\pm} =\omega_{\pm}(n,x)$ for $X(t)=x$ and $N(t)=n$. In between jumps in the discrete variable, $X(t)$ evolves according to the SDE
\begin{equation}
\label{pd}
dX=A(n,x)dt+\sqrt{2D(n,x)}dW
\end{equation}
for $N(t)=n$.
Introduce the probability density 
\begin{eqnarray}
 &P(n,x,t)dx=\mbox{Prob}\{X(t)\in (x,x+dx), N(t)=n\},
\end{eqnarray}
given an initial state $X(0)=x_0,N(0)=n_0$.
The probability density evolves according to the differential CK equation
\numparts
\begin{eqnarray}
\label{CK}
\fl &\frac{\partial P(n,x,t)}{\partial t}=-\frac{\partial A(n,x)P(n,x,t)}{\partial x}+\frac{\partial^2 D(n,x)P(n,x,t)}{\partial x^2}+\sum_{m\geq 0}Q_{nm}(x)P(m,x,t),\nonumber \\
\fl \\
\fl &\sum_{m\geq 0}Q_{nm}(x)P(m,x,t)=\omega_+(n,x-1)P(n,x-1,t)
\\ 
\fl  &\hspace{2cm} 
+  \omega_-(n,x+1)P(n,x+1,t) -[\omega_+(n,x)+\omega_-(n,x)]P(n,x,t).\nonumber
\end{eqnarray}
\endnumparts

Consider the Hilbert space spanned by the vectors $|n,x\rangle$ and carry over the definitions of the operator pairs $a,a^{\dagger}$ and $\hat{x},\hat{p}$ as follows\footnote{More precisely, we have the tensor product $|n,x\rangle=|n\rangle \otimes |x\rangle$ and the annihilation and creation operators take the form $(1,a)$ and $(1,a^{\dagger})$, respectively. Similarly the space and momentum operators are written as $(\hat{x},1)$ and $(\hat{p},1)$.}:
\numparts
\begin{eqnarray}
& a|n,x\rangle =n|n-1,x\rangle,\quad a^{\dagger}|n,x\rangle =|n+1,x\rangle,\\
& \hat{x}|n,x\rangle =x|n,x\rangle ,\quad \hat{p}|n,x\rangle = -i\overset{\leftarrow}{\frac{d}{d x}}|n,x\rangle.
\end{eqnarray}
\endnumparts
These preserve the commutation relations (\ref{com1}) and (\ref{com2}). We also have the completeness relation
\begin{equation}
\sum_{n\geq 0}\int_{-\infty}^{\infty} dx\, |n,x\rangle \langle n,x|=1.
\end{equation}
Introduce the state vector
\begin{equation}
|\psi(t)\rangle =\sum_{n\geq 0}\int_{-\infty}^{\infty} dx\,  P(n,x,t)|n,x\rangle.
\end{equation}
Differentiating both sides with respect to time gives
\begin{eqnarray*}
\fl \frac{d}{dt}|\psi(t)\rangle&= \sum_{n\geq 0}\int_{-\infty}^{\infty} dx\,\left [-\frac{\partial A(n,x)P(n,x,t)}{\partial x}+\frac{\partial^2 D(n,x)P(n,x,t)}{\partial x^2}\right ]|n,x\rangle\\
\fl &\hspace{3cm}+\sum_{n,m\geq 0}\int_{-\infty}^{\infty} dx\,Q_{nm}(x)P(m,x,t)|n,x\rangle\\
\fl &= \int_{-\infty}^{\infty} dx\,\left [-A(n,x)P(n,x,t)\overset{\leftarrow}{\frac{\partial }{\partial x}}+ D(n,x)P(n,x,t)\overset{\leftarrow}{\frac{\partial^2}{\partial x^2}}\right ]|n,x\rangle\\
\fl  &\hspace{3cm} +\hat{H}_{\rm bd}\sum_{n\geq 0}\int_{-\infty}^{\infty} dx\, P(n,x,t)|n,x\rangle\\
\fl &=\left [-i\hat{p}A(a^{\dagger}a,\hat{x})-\hat{p}^2 D(a^{\dagger}a,\hat{x})+\hat{H}_{\rm bd}\right ]\sum_{n\geq 0}\int_{-\infty}^{\infty} dx\, P(n,x,t)|n,x\rangle,
\end{eqnarray*}
where $\hat{H}_{\rm bd}$ is an extended version of equation (\ref{Hcal}):
\begin{equation}
\hat{H}_{\rm bd}=(a- a^{\dagger}a){\omega}_-(a^{\dagger}a,\hat{x})   +(a^{\dagger}-1)\omega_+(a^{\dagger}a,\hat{x}).
\end{equation}
Hence, as in the previous two examples, the CK equation can be written in the operator form
\numparts
\begin{equation}
\label{sh:ev}
\frac{d}{dt}|\psi(t)\rangle=\hat{H} |\psi(t)\rangle,
\end{equation}
with
\begin{equation}
\label{sh:H}
\hat{H}=-i\hat{p}A(a^{\dagger}a,\hat{x})-\hat{p}^2 D(a^{\dagger}a,\hat{x})+\hat{H}_{\rm bd}.
\end{equation}
\endnumparts
The formal solution of the FP equation is then
\begin{equation}
\label{sh:sol}
|\psi(t)\rangle =\e^{\hat{H}t}|\psi(0)\rangle ,
\end{equation}
and we can define expectations of the continuous and discrete variables according to
\numparts
\begin{eqnarray}
\langle X(t)\rangle &=\sum_{n\geq 0}\frac{1}{n!}\int_{-\infty}^{\infty} dx\, \langle n,x|\hat{x}\e^{\hat{H}t}|\psi(0)\rangle,\\
\langle N(t)\rangle &=\sum_{n\geq 0}\frac{1}{n!}\int_{-\infty}^{\infty} dx\, \langle n,x|a^{\dagger}a\e^{\hat{H}t}|\psi(0)\rangle .
\end{eqnarray}
\endnumparts

Rather than using the conjugate pair ($a,a^{\dagger}$) for the discrete component, we could treat $n/\overline{N}$ as a continuous variable $y$ and use the operators $(\hat{y},\hat{x})$. In that case, we set
\begin{equation}
\label{psi0}
|\psi(t)\rangle =\int_0^{\infty}dy\int_{-\infty}^{\infty} dx |y,x\rangle,
\end{equation}
and the operator $\hat{H}$ becomes
\begin{equation}
\label{psiH}
\hat{H}'=-i\hat{p}A(\hat{y},\hat{x})-\hat{p}^2 D(\hat{x},\hat{x})+\hat{H}'_{\rm bd},
\end{equation}
with $\hat{H}'_{\rm bd}$ given by equation (\ref{H'}).
Finally note that one could also consider other basis vectors such as $|n,p\rangle =|n\rangle \otimes |p\rangle$, where $|p\rangle$ is the momentum state vector, or $|\phi,x\rangle=|\phi\rangle \otimes |x\rangle$, where $|\phi\rangle$ is a coherent state.
Yet another example will be introduced in section 3 when we construct the stochastic hybrid path integral.

\section{Construction of stochastic hybrid path integral}
One of the advantages of expressing the evolution equation for the probability density  in terms of an operator equation acting on a Hilbert space, see equation (\ref{sh:ev}), is that it is relatively straightforward to construct a corresponding path integral representation of the solution. This has been carried out explicitly in the case of chemical master equations \cite{Doi76,Doi76a,Peliti85} and FP equations \cite{Holmes20}. Here we develop the analogous construction for the CK equation of a stochastic hybrid system. As with other stochastic processes, the first step is to divide the time interval $[0,t]$ into $N$ subintervals of size $\Delta t=t/N$ and rewrite the formal solution (\ref{sh:sol}) as
\begin{equation}
|\psi(t)\rangle =\e^{\hat{H}\Delta t}\e^{\hat{H}\Delta t}\cdots \e^{\hat{H}\Delta t}|\psi(0)\rangle,
\end{equation}
with $\hat{H}$ given by equation (\ref{sh:H}). We then insert multiple copies of an appropriately chosen completeness relation. In the case of a chemical master equation, one typically uses equation (\ref{bdcom2}) for coherent states, wheres for an FP equation one can use a combination of equations (\ref{fpcom1}) and (\ref{fpcom2}). Here we use the completeness relation for the basis vectors $|n,x\rangle$ so that
\begin{eqnarray}
& |\psi(t)\rangle = \sum_{n_0\geq 0}\frac{1}{n_0!}\cdots \sum_{n_N\geq 0}\frac{1}{n_N!}\int_{-\infty}^{\infty}dx_0\cdots  \int_{-\infty}^{\infty}dx_N
|n_N,x_N\rangle \nonumber\\
&\times \langle n_N,x_N|\e^{\hat{H}\Delta t}|n_{N-1},x_{N-1}\rangle \langle n_{N-1},x_{N-1}|\e^{\hat{H}\Delta t}|n_{N-2},x_{N-2}\rangle\nonumber \\
 &\cdots \times \langle n_1,x_1|\e^{\hat{H}\Delta t}|n_0,x_0\rangle \langle n_0,x_0|\psi(0)\rangle ,
 \label{pip}
\end{eqnarray}
with $ \langle n_0,x_0|\psi(0)\rangle=n_0!P(n_0,x_0,0)$.

In the limit $N\rightarrow \infty$ and $\Delta t \rightarrow 0$ with $N\Delta t =t$ fixed, we can make the approximation
\begin{eqnarray*}
\fl &\langle n_{j+1},x_{j+1}|\e^{\hat{H}\Delta t}| n_{j},x_{j}\rangle \approx \langle n_{j+1},x_{j+1}|1+\hat{H}\Delta t| n_{j},x_{j}\rangle=\delta(x_{j+1}-x_{j})n_j! \delta_{n_{j+1},n_{j}}\\
\fl &\hspace{0.5cm}+\Delta t\langle n_{j+1},x_{j+1} |\, \bigg (-i\hat{p}A(n_j,x_j)-\hat{p}^2 D(n_j,x_j) \bigg) \delta_{n_j,m_j}+\sum_{m_j\geq 0} Q_{m_jn_j}(x_j)\,  | m_{j},x_{j}  \rangle.
\end{eqnarray*}
Each small-time propagator thus has the matrix form (to first order in $\Delta t$)
\numparts
\begin{eqnarray}
\label{K1}
\fl &\langle n_{j+1},x_{j+1}|\e^{\hat{H}\Delta t}| n_{j},x_{j}\rangle\approx  \langle n_{j+1},x_{j+1}|1+\sum_{m_j}K_{m_jn_j}(x_j, \hat{p})| m_{j},x_{j}\rangle \Delta t,\\
\label{K}
\fl &K_{nm}(x,p)=[-ipA(n,x)-p^2 D(n,x)]\delta_{m,n} +Q_{nm}(x).
\end{eqnarray}
\endnumparts
Suppose that for fixed $x,p$, there exists a complete orthonormal set of right and left eigenvectors
$R_{\mu}(n,x,p)$ and $\overline{R}_{\mu}(n,x,p)$ such that
\numparts
\begin{eqnarray}
\label{eigR0}
\sum_{m\geq 0}K_{nm}(x,p)R_{\mu}(m,x,p)&=\Lambda_{\mu}(x,p)R_{\mu}(n,x,p),\\
\sum_{m\geq 0}K_{mn}(x,p)\overline{R}_{\mu}(m,x,p)&=\Lambda_{\mu}(x,p)\overline{R}_{\mu}(n,x,p),
\label{eigS0}
\end{eqnarray}
with
\begin{equation}
\label{RS}
\fl \sum_{m\geq 0}R_{\mu}(m,x,p)\overline{R}_{\nu}(m,x,p)=\delta_{\mu,\nu},\quad \sum_{\mu \geq 0}R_{\mu}(m,x,p)\overline{R}_{\mu}(n,x,p)=\delta_{m,n}.
\end{equation}
\endnumparts
For a given $x$ and $p$, we introduce the vectors
\begin{equation}
\label{Rmu}
|R_{\mu}(x,p)\rangle = \sum_{n\geq 0} R_{\mu}(n,x,p)|n,p\rangle,
\end{equation}
and their duals
\begin{equation}
\label{Smu}
\langle \overline{R}_{\mu}(x,p)| = \sum_{n\geq 0}\frac{1}{n!}\overline{R}_{\mu}(n,x,p)\langle n,p|.
\end{equation}
(We are assuming all functions are real. In the case of complex functions, the dual is defined in terms of the complex conjugate $\overline{R}_{\mu}^*$.) It immediately follows that
\begin{eqnarray*}
\fl\langle \overline{R}_{\nu}(x,q)|R_{\mu}(x,p)\rangle &= \sum_{n,m\geq 0}\frac{1}{n!} \overline{R}_{\nu}(n,x,q) R_{\mu}(m,x,p)\langle n,q|m,p\rangle\\
\fl &= 2\pi \sum_{n,m\geq 0} \overline{R}_{\nu}(n,x,p) R_{\mu}(m,x,p)\delta(p-q)\delta_{n,m}\\
\fl &=2\pi \delta(p-q) \sum_{n\geq 0} \overline{R}_{\nu}(n,x,p) R_{\mu}(n,x,p)=2\pi \delta(p-q)\delta_{\mu,\nu}.
\end{eqnarray*}
The associated completeness relation is
\begin{equation}
\label{shs:com}
\sum_{\mu\geq 0} \int_{-\infty}^{\infty}\frac{dp}{2\pi}\, |R_{\mu}(x,p)\rangle\langle \overline{R}_{\mu}(x,p)|=1.
\end{equation}
Note that the given set of basis vectors is a natural generalization of the eigenvectors of the matrix generator for a birth-death process, see equation (\ref{eig:bd}).

If we substitute the completeness relation (\ref{shs:com}) into the small-time propagator (\ref{K1}) we see that
\begin{eqnarray*}
\fl &\langle n_{j+1},x_{j+1}|\e^{\hat{H}\Delta t}| n_{j},x_{j}\rangle\\
\fl &\approx  \sum_{\mu_j\geq 0} \int_{-\infty}^{\infty}\frac{dp_j }{2\pi}\langle n_{j+1},x_{j+1}|R_{\mu_j}(x_j,p_j)\rangle  \langle \overline{R}_{\mu_j}(x_j,p_j)|1+\sum_{m_j}{K_{m_jn_j}(x_j,p_j)\Delta t}|m_{j},x_{j}\rangle.
\end{eqnarray*}
Evaluating the second inner product using equation (\ref{eigS0}) and the definition (\ref{Smu}) implies that
\begin{eqnarray*}
\fl &\langle \overline{R}_{\mu_j}(x_j,p_j) | 1+\sum_{m_j}{K_{m_jn_j}(x_j,p_j)\Delta t}|m_{j},x_{j}\rangle\\
\fl&=
 \sum_{n,m_j\geq 0}\frac{1}{n!}\overline{R}_{\mu_j}(n,x_j,p_j)\langle n,p_j|1+{K_{m_jn_j}(x_j,p_j)\Delta t}|m_{j},x_{j}\rangle\\
\fl  &=\e^{-ip_jx_j}\sum_{m_j}\overline{R}_{\mu_j}(m_j,x_j,p_j)(1+K_{m_jn_j}(x_j,p_j)\Delta t)\\
&=\e^{-ip_jx_j}(1+\Lambda_{\mu}(x_j,p_j)\Delta t)\overline{R}_{\mu_j}(n_j,x_j,p_j).
\end{eqnarray*}
Moreover, on using equation (\ref{Rmu}), the first inner product becomes
\begin{eqnarray*}
&\langle n_{j+1},x_{j+1}|R_{\mu_j}(x_j,p_j)\rangle=\langle n_{j+1},x_{j+1}| \sum_{n\geq 0} R_{\mu_j}(n,x_j,p_j)|n,p_j\rangle \\
&=(n_{j+1}! )\e^{ip_{j}x_{j+1}}R_{\mu_j}(n_{j+1} ,x_j,p_j).
\end{eqnarray*}
 Therefore,
\begin{eqnarray*}
\fl  &\langle n_{j+1},x_{j+1}|\e^{\hat{H}\Delta t}| n_{j},x_{j}\rangle\approx (n_{j+1}! ) \sum_{\mu_j\geq 0} \int_{-\infty}^{\infty}\frac{dp_j}{2\pi}\\
 \fl &\times  \e^{ip_{j}(x_{j+1}-x_j)}\e^{\Lambda_{\mu_j}(x_j,p_j)\Delta t}R_{\mu_j}(n_{j+1} ,x_j,p_j)\overline{R}_{\mu_j}(n_j,x_j,p_j)+O(\Delta t^2).\nonumber
\end{eqnarray*}

Substituting the expression for the small-time propagator back into equation (\ref{pip}) yields 
\begin{eqnarray}
\fl & P(n,x,t)=\frac{1}{n!}\langle n,x|\psi(t)\rangle \approx \sum_{n_0\geq 0} \cdots \sum_{n_N\geq 0} \sum_{\mu_0\geq 0} \cdots \sum_{\mu_{N-1}\geq 0}  \nonumber \\
\fl &\quad\times \int_{-\infty}^{\infty}\int_{-\infty}^{\infty}\frac{dx_0dp_0}{2\pi}\cdots  \int_{-\infty}^{\infty}\int_{-\infty}^{\infty}\frac{dx_{N}dp_{N}}{2\pi} \delta(x-x_N) \label{pip2}\\
\fl &\quad \times \prod_{j=0}^{N-1}\left [\e^{ip_{j}(x_{j+1}-x_j)}e^{\Lambda_{\mu_j}(x_j,p_j)\Delta t}R_{\mu_j}(n_{j+1} ,x_j,p_j)\overline{R}_{\mu_j}(n_j,x_j,p_j)\right ]P(n_0,x_0,0) . \nonumber
\end{eqnarray}
Finally, we take the continuum limit $N\rightarrow \infty,\Delta t\rightarrow 0$ with $N\Delta t=t$ fixed, $x_j=x(j\Delta t)$ and $p_j=p(j\Delta t)$. Exploiting the fact that the resulting path integral sums over continuous paths, it follows that
\begin{eqnarray*}
 &\sum_{n_{j+1}\geq 0}R_{\mu_{j}}(n_{j+1} ,x_{j},p_{j})\overline{R}_{\mu_{j+1}}(n_{j+1},x_{j+1},p_{j+1})\\
&\underset{\Delta \rightarrow 0}{\longrightarrow} \sum_{n_{j+1}\geq 0}R_{\mu_{j}}(n_{j+1} ,x_{j},p_{j})\overline{R}_{\mu_{j+1}}(n_{j+1},x_j,p_j)=\delta_{\mu_{j+1},\mu_j}
\end{eqnarray*}
for $j=0,1,\ldots N-1$. This enforces the condition $\mu_j=\mu$ for all $j$ in the product of exponentials so that in the limit $\Delta t \rightarrow 0$,
\begin{eqnarray*}
\fl&\prod_{j=0}^{N-1}\left [\e^{ip_{j}(x_{j+1}-x_j)}e^{\Lambda_{\mu}(x_j,p_j)\Delta t}\right ]=
\exp\left (\sum_{j=0}^{N-1} ip_{j}(x_{j+1}-x_j)+\Lambda_{\mu}(x_j,p_j)\Delta t\right )\\
\fl &\rightarrow \exp\left ({\int_{0}^{t}[ip\dot{x}+\Lambda_{\mu}(x,p)]d\tau}\right ).
\end{eqnarray*}
After performing a Wick rotation of the momentum variable, $p\rightarrow -ip$, we obtain the path integral representation
\numparts
\begin{eqnarray}
\fl & P(n,x,t)=\sum_{n_0\geq 0}\int_{-\infty}^{\infty}dx_0 G(n,x,t|n_0,x_0)P(n_0,x_0,0),
\label{piMs}\\ \label{G}
\fl & G(n,x,t|n_0,x_0)\\
\fl &=\iint \limits_{x(0)=x_0}^{x(t)=x}  {\mathcal D}[p]{\mathcal D}[x]\sum_{\mu\geq 0} R_{\mu}(n,x,p(t))\exp\left (-{\int_{0}^{t}[p\dot{x}-\Lambda_{\mu}(x,p)]d\tau}\right ) \overline{R}_{\mu}(n_0,x_{0},p(0)).\nonumber 
\end{eqnarray}
\endnumparts
Under the Wick rotation,
\begin{equation}
\label{K2}
K_{nm}(x,p)=[pA(n,x)+p^2 D(n,x)]\delta_{m,n} +Q_{nm}(x),
\end{equation}
and the eigenvalue equations (\ref{eigR0}) and (\ref{eigS0}) become
\numparts
\begin{eqnarray}
\label{eigR}
\fl &\sum_{m\geq 0}\left \{ [pA(n,x)+p^2 D(n,x)]\delta_{m,n} +Q_{nm}(x)\right \}R_{\mu}(m,x,p) =\Lambda_{\mu}(x,p)R_{\mu}(n,x,p)\nonumber ,\\ \fl   \\
\fl &\sum_{m\geq 0}\overline{R}_{\mu}(m,x,p)\left \{[pA(n,x)+p^2 D(n,x)]\delta_{m,n} +Q_{mn}(x)\right \}=\Lambda_{\mu}(x,p)\overline{R}_{\mu}(n,x,p).\nonumber \\
\fl
\label{eigS}
\end{eqnarray} 
\endnumparts

\noindent{\bf Some remarks}
\medskip

\noindent (i) Equations (\ref{piMs}) and (\ref{G}) with $D \equiv 0$ recover the path integral expression for PDMPs derived previously using an integral representation of the Dirac delta function \cite{Bressloff14,Bressloff15}. However, the new derivation based on operator methods has several advantages. (i) The eigenvalue equations (\ref{eigR}) and (\ref{eigS}) arise naturally within the bra-ket formulation due to the structure of the operator $\hat{H}$; their introduction in Ref. \cite{Bressloff14}) was rather ad-hoc. (ii) It extends the bra-ket approach to stochastic processes that was first introduced for master equations \cite{Doi76,Doi76a,Peliti85} and more recently developed for Fokker-Planck equations \cite{Holmes20}. (iii) The operator formalism provides a more flexible framework for including additional  factors such as Gaussian noise in the piecewise dynamics (\ref{pd}). 
\medskip

\noindent (ii) If $A$ and $D$ are independent of the discrete state $n$, then the eigenvalue equations (\ref{eigR}) and (\ref{eigS}) become
\begin{equation}
\fl \sum_{m\geq 0}Q_{nm}(x)R_{\mu}(m,x,p)=[\Lambda_{\mu}(x,p)-pA(x)-p^2D(x)]R_{\mu}(n,x,p).
\end{equation}
Let $r_{\mu}(m,x)$ be an eigenvector of the matrix generator ${\bf Q}(x)$ for fixed $x$, see equation (\ref{Qr}). That is $\sum_{m\geq 0}Q_{nm}(x)r_{\mu}(m,x)=\lambda_{\mu}(x)r_{\mu}(m,x)$. Then $R_{\mu}(m,x,p)=c(p)r_{\mu}(m,x)$ for arbitrary $c(p)$ and
\begin{equation}
\Lambda_{\mu}(x,p)=pA(x)+p^2D(x)+\lambda_{\mu}(x).
\end{equation}
The path integral (\ref{G}) reduces to the form
\begin{eqnarray}
\fl & G(n,x,t|n_0,x_0)=\iint \limits_{x(0)=x_0}^{x(t)=x}  {\mathcal D}[p]{\mathcal D}[x]\exp\left (-{\int_{0}^{t}[p\dot{x}-pA(x)-p^2D(x)]d\tau}\right )\\
\fl &\hspace{3cm} \times \sum_{\mu\geq 0} r_{\mu}(n,x)  \overline{r}_{\mu}(n_0,x_0)\exp\left ({\int_{0}^{t}\lambda_{\mu}(x)d\tau}\right ) .\nonumber
\end{eqnarray}
That is, the path integral of the continuous stochastic process decouples from the discrete process and we recover the standard action of a one-dimensional Ito SDE \cite{Martin73,Dom76,Janssen76,Holmes20}:
\[S[x,p]=\int_{0}^{t}[p\dot{x}-pA(x)-p^2D(x)]d\tau.\]
It also follows that if the initial state of the system is $P(n_0,x_0,0)=\rho_{n_0}(\bar{x})\delta (x_0-\bar{x})$, say, then 
\[\sum_{n_0\geq 0}\int_{-\infty}^{\infty}dx_0 \overline{r}_{\mu}(n_0,x_0)P(n_0,x_0,0)=\delta_{\mu,0},\] and the sum over $\mu$ in the path integral (\ref{G}) is restricted to $\mu=0$. On the other hand, when the continuous and discrete processes are mutually coupled, the resulting dynamics mixes the eigenstates of the Markov chain. In that case, a further approximation is needed in order to restrict the sum over $\mu$, see sections 4 and 5.
\medskip

\noindent (iii) The derivation carries over to higher-dimensional stochastic hybrid systems with $M$ continuous variables $x_{\alpha}$, $\alpha=1,\ldots M$. The  Ito SDE becomes
\begin{equation}
\label{pdmulti}
dX_{\alpha}=A_{\alpha}(n,\x)dt+\sqrt{2D_{\alpha}(n,\x)}dW_{\alpha}
\end{equation}
for $N(t)=n$, where $W_{\alpha}(t)$ are independent Wiener processes. The multivariate CK equation takes the form
\begin{eqnarray}
 \frac{\partial P}{\partial t}&=\sum_{{\alpha}=1}^M\left [-\frac{\partial}{\partial x_{\alpha}}(A_{\alpha}(n,\x)P(n,\x,t)) +\frac{\partial^2}{\partial x_{\alpha}^2}(D_{\alpha}(n,\x)P(n,\x,t))\right ]\nonumber
 \\ &+\sum_{m}Q(n,m;\x)P(m,\x,t).
\label{CK0}
\end{eqnarray}
Following along identical lines to the one-dimensional case, one obtains
a path-integral representation of the solution to equation (\ref{CK0}):
\begin{eqnarray}
\fl & P(n,\x,t)=\sum_{n_0\geq 0}\int_{-\infty}^{\infty}d\x_0 G(n,\x,t|n_0,\x_0)P(n_0,\x_0,0),
\end{eqnarray}
with
\begin{eqnarray}
\label{piMd}
\fl & G(n,\x,t|n_0,\x_0)\\
\fl &=\iint \limits_{\x(0)=\x_0}^{\x(t)=\x}  {\mathcal D}[\p]{\mathcal D}[\x]\sum_{\mu\geq 0} R_{\mu}(n,\x,\p(t))\exp\left (- S_{\mu}[\x,\p]\right )\overline{R}_{\mu}(n_0,\x_{0},\p(0)),\nonumber 
\end{eqnarray}
and
 \begin{equation}
S_{\mu}[\x,\p]=\int_0^{t}\left [\sum_{\alpha=1}^Mp_{\alpha}\dot{x}_{\alpha}-\Lambda_{\mu}(\x,\p)\right ]d\tau .
\label{actM}
\end{equation}
Here $\Lambda_{\mu}$ is an eigenvalue of the linear equation
\begin{eqnarray}
&\left [\sum_{m}Q_{nm}(\x) +\sum_{\alpha=1}^M(p_{\alpha} A_{\alpha}(n,\x)+p_{\alpha}^2 D_{\alpha}(n,\x)) \delta_{n,m} \right ]{R}_{\mu}(m,\x,\p) \nonumber\\ 
 &\quad =\Lambda_{\mu}(\x,\p) {R}_{\mu}(n,\x,\p).
\label{wow2}
\end{eqnarray}

\medskip

\noindent (iv) Suppose that we had started with the state vector representation (\ref{psi0}) and the associated operator $\hat{H}_{\rm bd}'$ of equation (\ref{psiH}). Inserting multiple copies of the completeness relations (\ref{fpcom1}), (\ref{fpcom2}), (\ref{zcom0}) and (\ref{zcom}) one can derive a path integral of the form
\begin{eqnarray}
\fl |\psi(t)\rangle \sim \iint {\mathcal D}[p]{\mathcal D}[x] \iint  {\mathcal D}[z]{\mathcal D}[q] \exp\left (-{\int_{0}^{t}[p\dot{x}+q\dot{z}-H(x,p,z,q)]d\tau}\right ) , 
\end{eqnarray}
with
\begin{equation}
\fl H=pA(z,x)+p^2 D(z,x)]  +(\e^{-q}-1)z\omega_-(z,x)+(\e^{q}-1)\omega_+(z,x).
\end{equation}
It is clear that this is considerably more complicated than the path integral given by equation (\ref{piMs}), since it involves the doubling up of the integration variables. On the other hand, it avoids the need to solve a possibly infinite-dimensional eigenvalue equation.

\section{Finite discrete systems and the Perron-Frobenius theorem}

Although we took the discrete part to evolve according to a birth-death process with associated operator $\hat{H}_{\rm bd}$, the above derivation holds for any master equation with matrix generator ${\bf Q}$, provided  that there exists a complete orthonormal set of right and left eigenvectors
$R_{\mu}(x,p,n)$ and $\overline{R}_{\mu}(x,p,n)$ satisfying equations (\ref{eigR}) and (\ref{eigS}). This may not hold if the discrete system is infinite-dimensional. On the other hand, if the underlying discrete space is finite-dimensional, then one can make stronger statements about the spectrum using the Perron-Frobenius theorem. We first recall some results for finite continuous-time Markov chains \cite{Grimmett}. Let $N(t)$, $0\leq N(t)\leq N_{\max}$, be a discrete random variable whose associated master equation has the general form
\begin{equation}
\frac{dP(m,t)}{dt}=\sum_{m=1}^{N_{\max}} W_{mn}P(n,t)-\left (\sum_{n=1}^{N_{\max} }W_{nm}\right )P(m,t),
\end{equation}
with $W_{mm}=0$ for all $m$. The corresponding matrix generator is related to the transition matrix ${\bf W}$ according to
\begin{equation}
Q_{mn}=W_{mn}-\delta_{n,m}\sum_{k}W_{kn}.
\end{equation}
The discrete process is said to be irreducible if there exists a $t>0$ such that $\e^{{\bf Q}t}>0$. This implies that any two states of the Markov chain can be connected in a finite time. Under such a condition, one can apply the Perron-Frobenius theorem for finite square matrices \cite{Grimmett}.

\begin{theorem}[Perron-Frobenius for positive square matrices]
Let ${\bf A}=(a_{ij})$ be an $n\times n$ positive matrix: $a_{ij}>0$ for all $1\le i,j \leq n$. Then the following results hold.

\begin{enumerate}
\item There exists a simple, positive eigenvalue $\lambda_{1}$ of ${\bf A}$. Hence, the left and right eigenspaces associated with $\lambda_1$ are one-dimensional.

\item The remaining (possibly complex) eigenvalues $\lambda_{2},\ldots,\lambda_{n}$ satisfy $|\lambda_{j}|<\lambda_1$.

\item The components of the eigenvector ${\bf v}$, ${\bf A}\v=\lambda_1\v$, are all positive, that is, $v_j>0$ for all $j=1,\ldots,n$. All other eigenvectors have at least one negative component.

\end{enumerate}
\end{theorem} 

\noindent In the case of the generator of an irreducible Markov chain, we have $\sum_{n}Q_{nm}=0$, which implies $\psi=(1,1,\ldots,1)$ is a left eigenvector of ${\bf Q}$ whose eigenvalue is zero. This corresponds to the principal or Perron eigenvalue, which means that there exists a corresponding probability distribution $\rho(n)$ (positive eigenvector) for which $\sum_{m}Q_{nm}\rho(m)=0$. We can identify $\rho$ as the unique stationary density. Moreover, the Perron Frobenius theorem ensures that all other eigenvalues have negative real parts, ensuring that the distribution $P(m,t)\rightarrow \rho(m)$ as $t\rightarrow \infty$.

One well-known method for constructing $\rho(n)$ is to note that in steady-state the master equation (\ref{bdbd}) satisfies $J(n)=J(n+1)$ with 
\[ J(n)=\omega_-(n)\rho(n)-\omega_+(n-1)\rho(n-1).\]
Using the fact that $n$ is a nonnegative integer, that is, $\rho(n)=0$ for $n<0$, it follows that $J(n)=0$ for all $n$. Hence, by iteration,
\begin{equation}
\label{ssd}
\rho(n)=\rho(0)\prod_{m=1}^n\frac{\omega_+(m-1)}{\omega_-(m)},
\end{equation}
with
\[\rho(0)=\left (1+\sum_{n=1}^{N_{\max}}\prod_{m=1}^n\frac{\omega_+(m-1)}{\omega_-(m)}\right )^{-1}.\]
For finite $N_{\max}$, such a solution exists provided that all birth/death rates are positive definite, which ensures irreducibility.

Returning to the stochastic hybrid system (\ref{pd}), suppose that the discrete component is finite-dimensional and $Q_{nm}(x)$ for any fixed $x$ is the generator of an irreducible Markov chain. For fixed $(x,p)$, introduce the modified matrix
\begin{equation}
\widehat{K}_{nm}(x,p)=K_{nm}(x,p)+\Gamma \delta_{n,m},
\end{equation}
with
\begin{equation}
\Gamma =\max_n 
\left \{[pA(n,x)+p^2 D(n,x)]\delta_{m,n} +Q_{nn}(x)\right \}.
\end{equation}
$\widehat{K}(x,p)$ is then an irreducible matrix and we can apply the Perron-Frobenius theorem. This establishes that ${\bf K}(x,p)$ for fixed $(x,p)$ has a unique positive right eigenvector, which we identify with $R_0$, and $\Lambda_0$ is the corresponding principal eigenvalue such that $\Lambda_0>\mbox{Re}(\Lambda_{\mu})$ for all $\mu=1,\ldots,N_{\max}$.

One major application of path integrals is determining the ``least action'' path(s) in the weak noise limit, which provides the dominant contribution to the rate of escape from a metastable state \cite{Bressloff14a}. In the case of the piecewise SDE (\ref{pd}), we define the weak noise limit by introducing the scalings ${\bf Q}\rightarrow {\bf Q}/\epsilon$ and $D\rightarrow \epsilon D$. The former represents fast switching between the discrete states (adiabatic limit), whereas the latter represents weak Gaussian noise. The presence of a spectral gap vis-a-vis the Perron Frobenius theorem means that we can restrict the summation over $\mu$ in equation (\ref{piMs}) to $\mu=0$. If we also scale the momentum variable according to $p\rightarrow p/\epsilon$, then we obtain the reduced path integral
\numparts
\begin{eqnarray}
\label{piMs2}
\fl & G(n,x,t|n_0,x_0)=\iint \limits_{x(0)=x_0}^{x(t)=x}  {\mathcal D}[p]{\mathcal D}[x] R_{0}(n,x,p(t))\e^{- {S[x,p]}/{\epsilon}}\overline{R}_{0}(n_0,x_{0},p(0)),
\end{eqnarray}
where $S$ is the action
\begin{equation}
S[x,p]=\int_0^{t}\left [p\dot{x}-\Lambda_0(x,p)\right ]d\tau .
\label{actM0}
\end{equation}
\endnumparts
Finding the least action path reduces to a classical variational problem in which the principal eigenvalue $\Lambda_0(x, p)$ acts as an effective Hamiltonian. In particular, the least action path is a solution of Hamilton's equations
\begin{equation}
\frac{dx}{dt}=\frac{\partial \Lambda_0}{\partial p},\quad \frac{dp}{dt}=-\frac{\partial \Lambda_0}{\partial x},
\end{equation}
for appropriately defined initial conditions. Equivalently, the action satisfies the Hamilton-Jacobi equation
\begin{equation}
\label{HJ}
\Lambda_0(x,\partial_xS)=0.
\end{equation}
The corresponding rate of escape from a metastable state is $\sim \e^{-S/\epsilon}$. Hence, solving an escape problem in the weak noise limit only requires determining the principal eigenvalue $\Lambda_0$ and the associated left and right eigenvectors $R_0,\overline{R}_0$. In the case of small $N_{\max}$ such as a two-state Markov chain, these can be calculated by brute force. However, there are only a few examples where it has been possible to calculate $\Lambda_0$ explicitly for arbitrarily large $N_{\max}$ \cite{Bressloff14a}, namely, when the steady-state distribution of the underlying birth-death process is a binomial or a Poisson distribution. Here we extend these calculations to include the effects of Gaussian noise in the $x$-dynamics.

\subsection{Binomial distribution} As our first example, consider the stochastic hybrid system (\ref{pd}) with 
\begin{equation}
\label{ion1}
dX=\left [\frac{n}{N_{\max}}f(X)-g(X)\right ]+\sqrt{\epsilon nD_0} dW,
\end{equation}
and switching governed by a birth-death process with transition rates 
\begin{equation}
\label{ion2}
\omega_+(n,x)=(N_{\max}-n)\alpha(x),\quad \omega_-(n,x)=\beta(x).
\end{equation}
For the sake of illustration, we assume that the strength of the Gaussian noise is proportional to $n$.
 The associated birth-death process could be a model for a population of $N_{\max}$ identical and independent two-state ion channels, each of which stochastically switches between an open and closed state at rates that depend on the continuous variable $X$. The discrete random variable $N(t)$ would represent the number of open ion channels at time $t$. For appropriate choices of the functions $f,g,\alpha,\beta$, this system can be interpreted as a model of voltage-dependent sodium or calcium ion channels \cite{Keener11,NBK13,Newby14}.
If $X(t)=x$ for all $t$, then equation (\ref{ssd}) implies that the stationary distribution of the birth-death process is the binomial distribution
\begin{equation}
\label{bluepp}
\fl \rho(n,x)=a(x)^n(1-a(x))^{N_{\max}-n}\frac{N_{\max}!}{n!(N_{\max}-n)!} ,\quad a(x)=\frac{\alpha(x)}{\alpha(x)+\beta(x)}.
\end{equation}
This can be rewritten in the more suggestive form
\[\rho(n,x)={\mathcal N}(x) \frac{\Gamma^n(x)}{(N_{\max}-n)!n!}, \ \Gamma(x)=\frac{\alpha(x)}{\beta(x)},
\]
with ${\mathcal N}(x)=N_{\max}!(1-a(x))^{N_{\max}}$ a normalization factor. 
It turns out that we can determine the principal eigenvalue of equation (\ref{Qr}) with $A_n,\omega_{\pm}(n,x)$ given by equations (\ref{ion1}) and (\ref{ion2}) by considering the positive trial solution \cite{Bressloff14}
\begin{equation}
R_0(n,x,p)=\frac{\Gamma^n(x,p)}{(N_{\max}-n)!n!}.
\end{equation}
Substituting into equation (\ref{Qr}) 
yields the following equation relating $\Gamma$ and $\Lambda_0$:
\begin{eqnarray*}
\fl p\left (\frac{n}{N_{\max}}f-g\right )+p^2D_0n+\frac{n\alpha}{\Gamma}+\Gamma \beta(N_{\max}-n)   -n\beta-(N_{\max}-n)\alpha =\Lambda_0 .
\end{eqnarray*}
Collecting terms independent of $n$ and terms linear in $n$ yields a pair of equations
for $\Gamma$ and $\Lambda_0$:
\begin{equation}
\label{pp}
p\frac{f(x)}{N_{\max}}+p^2D_0=- \left (\frac{1}{\Gamma(x,p)}+1 \right )\left (\alpha(x)-\beta(x) \Gamma(x,p) \right ),
\end{equation}
and
\begin{equation}
\label{ll}
\Lambda_0(x,p)=-N(\alpha(x)-\Gamma(x,p) \beta(x))-pg(x).
\end{equation}
 Eliminating $\Gamma$ then yields 
a quadratic equation for $\Lambda_0$ of the form
\begin{equation}
\label{ll0}
\Lambda_0^2+\sigma(x,p)\Lambda_0-h(x,p)=0,
\end{equation}
with
\begin{eqnarray*}
\sigma(x,p)&=&p(2g(x)-f(x)-pD_0)+N_{\max}(\alpha(x)+\beta(x)),\\ 
h(x,p)&=&p\bigg  [g(x)\bigg (p(f(x)-g(x))+p^2D_0-N_{\max}(\alpha(x)+\beta(x)\bigg ) \\
&&\quad +N_{\max}\alpha(x)(f(x)+pD_0N_{\max})\bigg ].
\end{eqnarray*}
It follows that the Hamilton-Jacobi equation (\ref{HJ}) becomes $h(x,\partial_xS)=0$. Solving the latter equation determines the leading order exponential contribution to the mean time to escape from a metastable state \cite{NBK13,Newby14}. Differentiating the quadratic equation (\ref{ll0}) and setting $\Lambda_0=0$ also determines Hamilton's equations:
\begin{eqnarray}
\dot{x}=\frac{\partial \Lambda_0}{\partial p}=\frac{1}{\sigma(x,p)}\frac{\partial h}{\partial p},\quad \dot{p}=-\frac{\partial \Lambda_0}{\partial x}=-\frac{1}{\sigma(x,p)}\frac{\partial h}{\partial x}.
\end{eqnarray}
In particular, one solution is $p=0$ and
\begin{equation}
\dot{x}=\frac{1}{\sigma(x,0)}\left .\frac{\partial h}{\partial p}\right |_{p=0}=\frac{\alpha(x)}{\alpha(x)+\beta(x)}f(x)-g(x).
\end{equation}
This is precisely the deterministic rate equation that holds in the zero noise limit $\epsilon\rightarrow 0$.

\subsection{Poisson distribution}
We now consider an example for which a positive eigenfunction can be constructed even though the Perron-Frobenius theorem does not strictly apply. Take $N(t)\in \Z^+$ with transition rates
\begin{equation}
\label{nn1}
\omega_+(n)=f(X),\quad \omega_-(n)=1.
\end{equation}
The continuous variable evolves according to the piecewise SDE
\begin{equation}
\label{nn2}
dX=(-X+n)dt+\sqrt{\epsilon nD_0}dW
\end{equation}
for $N(t)=n$. We assume that $f$ is a bounded positive function. One application of this type of PDMP is to a one-population neural network model \cite{Bressloff13a,Bressloff15}. If $X(t)=x$ is fixed for all $t$, then the steady-state distribution of the birth-death process is determined from equation (\ref{ssd}), and is given by a Poisson distribution:
\begin{equation}
\rho(n,x)=\frac{f(x)^n}{n!}\e^{-f(x)}.
\end{equation}
The eigenvalue equation (\ref{eigR}) becomes
\begin{eqnarray}
\fl &[p(-x+n) +p^2 D_0]R_{\mu}(n,x,p)+ f(x)R_{\mu}(n-1,x,p)+(n+1)R_{\mu}(n+1,x,p)\nonumber \\
\fl &\quad -(f(x)+n)R_{\mu}(n,x,p)=\Lambda_{\mu}(x,p)R_{\mu}(n,x,p).
\label{Per1}
\end{eqnarray}
In terms of a candidate principal eigenvalue and associated positive eigenvector, we can formally solve this equation using the trial positive solution
\begin{equation}
\label{Rlam}
R_0(n,x,p)=\frac{\Gamma^n(x,p)}{n!}.
\end{equation}
This  yields the following equation relating $\Lambda_0$ and $\Gamma$:
\begin{equation*}
\left [\frac{ f(x)}{\Gamma}-1\right ]n+\Gamma-f(x)+p(-x+n)+p^2D_0=\Lambda_0.
\end{equation*}
Collecting terms independent of $n$ and linear in $n$, we find that 
\begin{equation}
\Gamma=\frac{f(x)}{1-p-p^2D_0},\quad \Lambda_0=\frac{pf(x)}{1-p-p^2D_0}-px.
\label{Rlam2}
\end{equation}
The deterministic rate equation is recovered from Hamilton's equations when $p=0$:
\begin{equation}
\label{rate2}
\dot{x}=\left . \frac{\partial \Lambda_0}{\partial p} \right |_{p=0} =-x+f(x).
\end{equation}

In contrast to the previous example, the candidate principal eigenvalue has a pair of real singularities at 
\begin{equation}
p=p_{\pm}\equiv \frac{1}{2D_0}\left (-1\pm \sqrt{1+4D_0}\right ),\quad  D_0 >0.
\end{equation}
Moreover, $\Gamma(x,p)<0$ for $p>p_+$ and $p<p_-$, contradicting the requirement that the eigenfunction $R_0$ is positive. In the special case $D_0=0$, there is a single pole at $p=1$ and $\Gamma(x,p)<0$ for $p>1$.
The origin of the singularity can be understood by considering a large but finite population size $N_{\max}$. The Perron-Frobenius theorem then holds but the solution of the eigenvalue equation becomes non-trivial. The basic difficulty arises because the above ansatz for $R_0$ does not satisfy the boundary condition at $n=N_{\max}$. For the sake of illustration suppose that $D_0=0$. Setting $n=N_{\max}$, $\mu=0$ and $R_{0}(N_{\max}+1,x,p)=0$ in equation (\ref{Per1}), we have
\begin{eqnarray*}
 \fl &p(-x+N_{\max}) R_{0}(N_{\max},x,p)+ f(x)R_{0}(N_{\max}-1,x,p) -(f(x)+N_{\max})R_{0}(N_{\max},x,p)\\
\fl  & \quad =\Lambda_{0}(x,p)R_{0}(N_{\max},x,p),
\end{eqnarray*}
which can be rearranged to give
\[R_{0}(N_{\max},x,p)=\frac{f(x)R_{0}(N_{\max}-1,x,p)}{\Lambda_0(x,p)+f(x)+px+N_{\max}(1-p)}.\]
This clearly does not satisfy the ansatz (\ref{Rlam}). However, if $p<1$ then in the large-$N_{\max}$ limit, we set that
\[\fl R_{0}(N_{\max},x,p)\rightarrow \frac{f(x)}{N_{\max}(1-p)}R_{0}(N_{\max}-1,x,p)=\frac{\Gamma(x,p)}{N_{\max}}R_{0}(N_{\max}-1,x,p).\]
This shows that the given ansatz is a good approximation to the eigenvector for large $N_{\max}$ and $p<1$. Keeping $N_{\max}$ large but finite means that we can ensure a spectral gap using Perron-Frobenius.

\subsection{Variational method} 

For a general stochastic hybrid system it is not possible to construct an explicit solution for the principal eigenvalue. However, one can estimate the principal eigenvalue (assuming it exists) by using a Ritz variational method. (The latter is often used to estimate the ground state energy of a quantum mechanical system.) Suppose that $(x,p)$ is fixed and consider the general eigenfunction expansion
\begin{equation}
V(n)=\sum_{\mu\geq 0}v_{\mu}R_{\mu}(n).
\end{equation}
We have suppressed the dependence on $(x,p)$ and are assuming that the eigenvectors $R_{\mu}$ form a complete orthonormal basis.
Define the so-called Rayleigh quotient
\[\Upsilon[V]=\frac{\langle V,{\bf K}V\rangle }{\langle V,V\rangle},
\]
where ${\bf K}$ is the matrix (\ref{K2}).
Substituting for $V$ using the eigenfunction expansion and exploiting orthonormality of the eigenfunctions shows that
\begin{eqnarray*}
\fl \langle V,{\mathbf K}V\rangle  &=\sum_{n,m\geq 0}\left (\sum_{\mu\geq 0} v_{\mu}R_{\mu}(n)\right ) K_{nm}\left (\sum_{\nu} v_{\nu}R_{\nu}(m)\right )  \\
\fl &=\sum_{\mu,\nu}v_{\mu}v_{\mu} \sum_{n,m\geq 0}R_{\mu}(n) K_{nm}R_{\nu}(m)=\sum_{\mu,\nu}\Lambda_{\nu}v_{\mu}v_{\mu} \sum_{n\geq 0}R_{\mu}(n) R_{\nu}(n)\\
\fl &=\sum_{\mu,\nu}\Lambda_{\nu}v_{\mu}v_{\mu} \delta_{\nu,\mu}=\sum_{\mu }\Lambda_{\mu}|v_{\mu}|^2.
\end{eqnarray*}
Performing a similar calculation for the denominator of the Rayleigh coefficient thus yields
\begin{equation}
\Upsilon[V] = \frac{\sum_{\mu }\Lambda_{\mu}|v_{\mu}|^2}{\sum_{\mu }|v_{\mu}|^2}\leq \Lambda_{0}.
\end{equation}
The Ritz variational method consists of introducing a finite set 
of functions $\Psi_i$, $i=1,\ldots,M$, and considering the trial solution
$V=\sum_{i=1}^M v_i\Psi_i$. This is substituted into the Rayleigh quotient $\Upsilon[V]$, which is then maximized with respect to the coefficients $v_j$. The art of this method is choosing an appropriate set of trial vectors $\Psi_i$.

\section{Perturbation methods}

In addition to large deviation theory, stochastic path integrals provide a systematic method for performing various perturbation expansions in the weakly nonlinear or weak noise limits \cite{Zinn02,Kleinert09}. These are often supplemented by graphical representations of the various terms in the perturbation expansion analogous to Feynman diagrams in quantum theory \cite{Feynman48}. For a nice pedagogical review of such methods for SDEs, see \cite{Chow15}. 

\subsection{Small momentum expansion and the diffusion approximation}

The weak noise limit for stochastic path integrals is analogous to the semi-classical limit in quantum mechanics. In both cases one can obtain a Gaussian approximation of the path integral by performing a small momentum expansion. In order to apply this to the hybrid path integral (\ref{piMs2}), we need to determine the corresponding expansion of the principal eigenvalue $\Lambda_0$ 
and the associated eigenvector $R_0$. (In the following we drop the subscript $\mu=0$.) Therefore, introduce the rescaling $p\rightarrow \epsilon p$ and the series expansions
\[\fl R(m,x,\epsilon p)=R_0(m,x)+\epsilon p R_1(m,x)+\epsilon^2 p^2 R_2(m,x)+\ldots,\quad \sum_m R_k(m,x)=\delta_{k,0},\]
\[\Lambda(x,\epsilon p)=\Lambda_0(x)+\epsilon p \Lambda_1(x)+\epsilon^2 p^2 \Lambda_2(x)+\ldots .\]
 Substituting into the eigenvalue equation (\ref{eigR}),
 \begin{eqnarray*}
\fl & \sum_{m  }\left [A_{nm}(x) +  (\epsilon pF(m,x)+\epsilon^2p^2D(m,x)) \delta_{n,m}\right ]\bigg (R_0(m,x)+\epsilon p R_1(m,x)\\
\fl&\quad +\epsilon^2 p^2 R_2(m,x)+\ldots\bigg )=\left [ \Lambda_0(x)+\epsilon p \Lambda_1(x)+\epsilon^2 p^2 \Lambda_2(x)+\ldots\right ]\\
\fl &\hspace{4cm} \times \bigg (R_0(n,x)+\epsilon p R_1(n,x)+\epsilon^2 p^2 R_2(n,x)+\ldots\bigg ).
\end{eqnarray*}
Collecting terms in equal powers of $\epsilon$ yields a hierarchy of equations. The first three are
\begin{eqnarray*}
\fl &\sum_{m }[A_{nm}(x)-\Lambda_0(x)\delta_{n,m}]R_0(m,x)=0,\\
\fl &\sum_{m  }[A_{nm}(x)-\Lambda_0(x)\delta_{n,m}]R_1(m,x)=-F(n,x)R_0(n,x) +\Lambda_1(x)R_0(n,x),\\
\fl &\sum_{m  }[A_{nm}(x)-\Lambda_0(x)\delta_{n,m}]R_2(m,x) =-F(n,x)R_1(n,x)+\Lambda_1(x)R_1(n,x)\\
\fl &\hspace{6cm}+ (\Lambda_2(x)-D(n,x))R_0(n,x).
\end{eqnarray*}
The first equation has the solution $\Lambda_0(x)=0$ and $R_0(m,x)=\rho(m,x)$.
Applying the Fredholm alternative theorem to the second and third equations by summing over $n$ gives the self-consistency conditions
\begin{eqnarray*}
\fl 0&=\sum_n\left \{-F(n,x)R_0(n,x) +\Lambda_1(x)R_0(n,x)\right \},\\
\fl 0&=\sum_n\left \{-F(n,x)R_1(n,x)+\Lambda_1(x)R_1(n,x) + [\Lambda_2(x)-D(n,x)]R_0(n,x)\right \}.
\end{eqnarray*}
The normalization conditions $\sum_m R_k(m,x)=\delta_{k,0}$ imply that
\begin{eqnarray*}
\Lambda_1(x)&=\sum_nF(n,x)\rho(n,x),\\
\Lambda_2(x)&=\sum_n \left (F(n,x) R_1(n,x)+D(n,x) R_0(n,x)\right ).
\end{eqnarray*}

Ignoring higher-order terms, we thus have the following Gaussian approximation of the principal eigenvalue:
\begin{equation}
\Lambda(x,p)\approx \epsilon p \overline{F}(x)+\epsilon^2 p^2\left (\sum_n F(n,x) Z(n,x)+\overline{D}(x)\right ),
\end{equation}
with 
\begin{equation}
\overline{F}(x)=\sum_nF(n,x)\rho(n,x),\quad \overline{D}(x)=\sum_nD(n,x)\rho(n,x),
\end{equation}
and $Z(n)$ satisfies the liner equation
\begin{equation}
\label{Z}
\sum_{m } A_{nm}(x)Z(m,x)=[\overline{F}(x)-F(n,x)]\rho(n,x),\ \sum_nZ(n,x)=0.
\end{equation}
Note that a unique solution for $Z(n,x)$ exists even though the matrix ${\bf A}(x)$ is singular, which is a consequence of the Fredholm alternative theorem. Substitute the Gaussian approximation into the path integral representation
\begin{equation*}
P(x,t|x_0,0)\sim \iint \limits_{x(0)=x_0}^{x(\tau)=x} D[x]D[p] \e^{-S[x,p]/\epsilon},
\end{equation*}
with the action
\begin{equation*}
S[x,p]=\int_0^{t}\left [\epsilon p(\dot{x}- \overline{F}(x))-\epsilon^2 p^2\left (\sum_n F(n,x) Z(n,x)+\overline{D}(x)\right )\right ]d\tau.
\end{equation*}
Performing the Gaussian integral with respect to $p$ after rotating in the complex plane results in the Onsager-Machlup path integral 
\begin{eqnarray} 
&P(x,t|x_0,0)\sim \iint \limits_{x(0)=x_0}^{x(t)=x} D[x] \exp\left (-\int_0^{t}\frac{[\dot{x}-\overline{F}(x)]^2}{4\epsilon D(x)}d\tau\right ),
\end{eqnarray}
with
\begin{equation}
\label{DD}
D(x)= \sum_n F(n,x) Z(n,x)+\overline{D}(x).
\end{equation}
This path integral is identical in form to the representation of solutions to the FP equation for the corresponding Ito SDE
\begin{equation}
\label{sde}
dX=\overline{F}(X)dt+\sqrt{2\epsilon D(X)}dW(t).
\end{equation}
This recovers the well-known quasi-steady-state approximation of a stochastic hybrid system \cite{Bressloff14,Bressloff14a}. 
Taking expectations of the SDE yields the deterministic rate equation under the Gaussian approximation
\begin{equation}
\label{rate}
\frac{d\langle X\rangle}{dt}=\overline{F}(\langle X\rangle ).
\end{equation}

\subsection{Moments equations and the loop expansion}

Another useful feature of path-integrals is that they provide a systematic method for generating a hierarchy of moment equations \cite{Zinn02,Kleinert09}. We will illustrate this by obtaining the leading order correction to the rate equation (\ref{rate}) that couples the first and second order moments. (A similar type of perturbation analysis has previously been applied to a neural master equation \cite{Buice10}.) The first step is to introduce the generating functional
\begin{equation}
\fl Z[{J},\widetilde{J}]
\sim \int  D[x]D[p]\exp \left (-\frac{S[x,p]}{\epsilon}+\frac{1}{\epsilon}\int_0^t \left [x(\tau)\widetilde{J}(\tau)+J(\tau)p(\tau)\right ]d\tau\right ).
\label{piZ}
\end{equation}
Various moments of physical interest can then be obtained by taking functional derivatives with respect to the ``current sources'' $J,\widetilde{J}$. For example,
\numparts
\begin{eqnarray}
\label{m1}
\langle   x(t)\rangle &=\epsilon  \left .  \frac{\delta}{\delta \widetilde{J}(t)}Z[{ J},\widetilde{ J}]\right |_{{ J}=\widetilde{ J}=0}\\
\label{m2}
\langle   x(t) x(t')\rangle  &=\epsilon^2  \left .  \frac{\delta}{\delta \widetilde{J} (t)}\frac{\delta}{\delta \widetilde{J}(t)}Z[{ J},\widetilde{ J}]\right |_{{ J}=\widetilde{ J}=0}\\
\label{m3}
\langle   x(t)p(t')\rangle   &=\epsilon^2 \left .  \frac{\delta}{\delta \widetilde{J}(t)}\frac{\delta}{\delta J(t)}Z[{ J},\widetilde{ J}]\right |_{{ J}=\widetilde{ J}=0}.
\end{eqnarray}
\endnumparts
Similarly, cumulants are obtained by functionally differentiating $W[J,\widetilde{J}]=-\epsilon \ln Z[J,\widetilde{J}[$.
The physical significance of the moment $\langle x(t)p(t')\rangle$ is in terms of the linear response of the system to an external input $h(t)$. 
That is, suppose a small external source term $h(t)$ is added to the right-hand side of the rate equation (\ref{rate}).
Linearizing about the solution $X_0=\langle X\rangle_{h=0}$ and setting $x=\langle X\rangle -X_0$ gives the non-autonomous linear equation
\[\frac{dx}{dt}=\overline{F}'(X_0(t))x+h(t).\]
Introducing the Green's function or propagator $G(t,t')$ according to the adjoint equation
\begin{eqnarray}
-\frac{d G(t,t')}{dt'}&=&\overline{F}'(X_0(t))x(t)+\delta(t-t'),
\end{eqnarray}
 the linear response is expressed as
\begin{equation}
x(t)=\int_0^t G(t,t') h(t')dt' .
\end{equation}
In other words, in terms of functional derivatives
\begin{equation}
 G(t,t')=\left . \frac{\delta \langle X(t)\rangle }{\delta h(t')}\right |_{h=0} =\langle   x(t)p(t')\rangle   .
\end{equation}

In the weak noise limit, we can apply a so-called loop expansion of the path-integral (\ref{piZ}), 
which is a diagrammatic method for carrying out an $\epsilon$ expansion based on steepest descents or the saddle--point method \cite{Zinn02,Kleinert09}. First, we introduce the exact means
\[\nu = \langle X\rangle  ,\quad \widetilde{\nu} =   \langle P \rangle,\]
where $P$ represents a stochastic momentum variable rather than a probability, and shift the variables by
\[x(t)\rightarrow  x(t)+\nu (t),\quad p (t)= p (t)+  \widetilde{\nu} (t).\]
Expanding the action in (\ref{piZ}) to second order in the shifted variables $ x, p$ yields an infinite--dimensional Gaussian integral, which can be formally evaluated to give
\begin{eqnarray*}
\fl Z[{ J},\widetilde{ J}]
&\sim&{\rm Det}\left [{\mathcal A}[{ \nu},\widetilde{ \nu}]\right ]^{-1/2}\exp\left (-\frac{S[{ \nu},\widetilde{ \nu}]}{\epsilon}+\frac{1}{\epsilon}\int \ dt  \left [ \nu (t)\widetilde{J} (t)+{J} (t)\widetilde{\nu} (t)\right ] \right ),
\end{eqnarray*}
where ${\mathcal A}[{ \nu},\widetilde{ \nu}]$ is the infinite-dimensional matrix with components
\begin{equation}
{\mathcal A}[{ \nu},\widetilde{ \nu}]_{r ,s }(t,t')=\left . \frac{\delta^2 S}{\delta u_ {r}(t)\delta u_ {s}(t')}\right |_{u_1= \nu,u_2=\widetilde{\nu}},\quad r,s=1,2,
\label{calD}
\end{equation}
and $u_1=x,u_2=p$.
Using the matrix identity ${\rm Det}{\bm M} =\e^{{\rm Tr}\log {\bm M}}$
we obtain the approximation
\begin{eqnarray}
Z[{ J},\widetilde{ J}]\sim  \e^{-S_{\rm eff}[{ \nu},\widetilde{ \nu}]/\epsilon}e^{\int \ dt \sum \left [ \nu (t)\widetilde{J} (t)+{J} (t)\widetilde{\nu} (t)\right ]/\epsilon },
\label{ZZ}
\end{eqnarray}
where
\begin{equation}
\label{Seff}
S_{\rm eff}[{ \nu},\widetilde{ \nu}]=S[{ \nu},\widetilde{ \nu}]+\frac{\epsilon}{2}{\rm Tr}\log \left [{\mathcal A}[{ \nu},\widetilde{ \nu}]\right ].
\end{equation}

In order to use the above expansion to derive moment equations, it is necessary to introduce a little more formalism. First, we extend the domain of definition of the variables $\nu,\widetilde{\nu}$ to the case of non-zero currents $J,\widetilde{J}$ according to
\begin{equation}
\nu(t)= -\frac{\delta W}{\delta  \widetilde{J}(t)},\quad \widetilde{\nu}(t)= -\frac{\delta W}{\delta{J}(t)}.
\label{Meff}
\end{equation}
The physical quantities are recovered by taking $J,\widetilde{J}\rightarrow 0$.
Now consider the Legendre transformation
\begin{equation}
\Gamma[ { \nu},\widetilde{ \nu}]= W[{ J},\widetilde{ J}]+\int \ dt \sum\left [{\nu}(t)\widetilde{J}(t)+{J}(t)\widetilde{\nu}(t)\right ],
\label{effact}
\end{equation}
where $\Gamma[{ \nu},\widetilde{ \nu}]= S_{\rm eff}[{ \nu},\widetilde{ \nu}] +{\mathcal O}(\epsilon^2)$ is known as the effective action. 
It follows from functionally differentiating equation (\ref{effact}) that
\begin{equation}
\widetilde{J}(t)= \frac{\delta \Gamma}{\delta {\nu}(t)},\quad {J}(t)= \frac{\delta \Gamma}{\delta \widetilde{\nu}(t)}.
\label{Jeff}
\end{equation}
Another useful result is obtained by functionally differentiating equations (\ref{Meff}) with respect to ${ \nu},\widetilde{ \nu}$:
\begin{eqnarray*}
\fl \delta(t-t')\delta_{r,s} &=\frac{\delta \nu_r(t)}{\delta \nu_s(t')} =-\frac{\delta^2 W}{\delta J_r(t)\delta \nu_s(t')}= -\sum_{q=1,2}  \int \frac{\delta^2 W}{\delta J_r(t)\delta J_q(\tau)}\frac{\delta J_q(\tau)}{\delta \nu_s(t')}d\tau,
\end{eqnarray*}
where $\nu_1=\nu,\nu_2=\widetilde{\nu}$ and $J_1=\widetilde{J},J_2=J$.
Differentiating equations (\ref{Jeff}) with respect to ${ J},\widetilde{ J}$ then shows that
\begin{eqnarray*}
&\sum_{q=1,2}  \int \frac{\delta^2 W}{\delta J_r(t)\delta J_q(\tau)}\frac{\delta^2 \Gamma}{\delta \nu_q(\tau)\delta \nu_s(t')}d\tau  =-\delta_{r,s} \delta(t-t').
\end{eqnarray*}
In other words, defining the infinite--dimensional matrix $\widehat{\mathcal A} [{ \nu},\widetilde{ \nu}]$ according to
\begin{equation}
\label{calD2}
\widehat{\mathcal A} [{ \nu},\widetilde{ \nu}]_{r,s}( t ,t')=\frac{\delta^2 \Gamma}{\delta \nu_r(t)\delta \nu_s(t')},
\end{equation}
we see that $\widehat{\mathcal A}[{ \nu},\widetilde{ \nu}]$ is the inverse of the two--point covariance matrix with components 
\begin{eqnarray*}
&C_{r ,s }(t,t')\equiv- \frac{\delta^2 W}{\delta J_r(t)\delta J_s(t')}=   \langle u_r(t) u_s(t')\rangle  - \langle u_r(t) \rangle \langle u_s(t')\rangle   .
 \end{eqnarray*}
Equations (\ref{calD}), (\ref{Seff}) and (\ref{calD2}) imply that $\widehat{\mathcal A}[{ \nu},\widetilde{ \nu}] = {\mathcal A}[{ \nu},\widetilde{ \nu}]+{\mathcal O}(\epsilon)$, that is, we can take ${\mathcal A}[{ \nu},\widetilde{ \nu}]$ to be the inverse of the two--point covariance matrix.

A dynamical equation for the physical mean $\nu(t)$ can be obtained by setting ${ J}=\widetilde{ J}=\widetilde{ \nu}=0$ in equations (\ref{Jeff}). In particular, 
\begin{eqnarray}
\fl 0=\left .\frac{\delta \Gamma[{ \nu},\widetilde{ \nu}]}{\delta{\widetilde{\nu}}(t)} \right |_{\widetilde{ \nu}=0}&=&\left .\frac{\delta S[x,p]}{\delta p(t)}\right |_{x={ \nu},p=0} +\left .\frac{\epsilon}{2}{\rm Tr}{\mathcal A}[x,p]^{-1}\frac{\delta {\mathcal A}[x,p]}{\delta p(t)} \right |_{x={ \nu},p=0},\nonumber
\label{trace}
\end{eqnarray}
with
\begin{eqnarray*}
\fl &\left . {\rm Tr}{\mathcal A}[x,p]^{-1}\frac{\delta {\mathcal A}[x,p]}{\delta p(t)} \right |_{x={ \nu},p=0}=\int dt' \int dt'' \sum_{r ,s } C_{r ,s }(t',t'')\left .\frac{\delta}{\delta p(t)}  \frac{\delta^2 S[x,p]}{\delta u_r(t')u_s(t'')}\right |_{x={ \nu},p=0}.
\end{eqnarray*}
The functional derivative in the above equation forces $t=t'=t''$. We now use the fact that the only non--vanishing, equal--time two--point correlation functions when $p=0$ is for $r=s=1$ \cite{Zinn02}. For example, the linear response $\langle x(t)p(t)\rangle=0$ is a consequence  of taking the noise to be Ito. It follows that 
\begin{eqnarray*}
&\left . {\rm Tr}{\mathcal A}[x,p]^{-1}\frac{\delta {\mathcal A}[x,p]}{\delta p(t)} \right |_{x={ \nu},p=0}= -\sigma (t)\left . \frac{\partial^3 \Lambda_0(x,p)}{\partial p\partial^2 x}\right |_{x={ \nu},p=0},
\end{eqnarray*}
where $\sigma(t)$ is the variance
\[\sigma (t)=\langle x^2(t) \rangle-\langle x(t) \rangle \langle x(t)\rangle  .\]
Hence, the leading-order correction to the deterministic rate equation (\ref{rate}) couples the mean $\nu(t)$ to the variance $\sigma(t)$:
\begin{equation}
\frac{d\nu}{dt}=\overline{F}(\nu)+\frac{\epsilon}{2}\sigma (t)\left . \frac{\partial^3 \Lambda_0(x,p)}{\partial p\partial^2 x}\right |_{x={ \nu},p=0}.
\end{equation}

It is also possible to extend the above loop expansion to calculate higher-order moment equations. However, the lowest $O(\epsilon)$ equation for $\sigma(t)$ can be obtained directly from equation (\ref{sde}) using a linear noise approximation. More specifically, set $Y(t)=(X(t)-\nu(t))/\sqrt{\epsilon}$ and linearize equation (\ref{sde}):
\begin{equation}
\label{sdelin}
dY=\overline{F}'(\nu)Ydt+\sqrt{2 D(X)}dW(t).
\end{equation}
We then have
\begin{eqnarray*}
\fl \langle Y(t+dt)Y(t+dt)\rangle &= \bigg \langle \left [Y(t)+\overline{F}'(\nu)Y(t)dt+\sqrt{2 D(\nu)}dW(t)\right ]^2\bigg \rangle\\
\fl &= \langle Y(t)Y(t)\rangle+2\overline{F}'(\nu)\langle Y(t)^2\rangle dt+2 D(\nu)\langle dW(t)^2\rangle+O(dt^2)\\
\fl &=\langle Y(t)Y(t)\rangle+\frac{2}{ \epsilon }\overline{F}'(\nu)\sigma(t)dt +2 D(\nu)dt+O(dt^2).
\end{eqnarray*}
Rearranging, dividing by $dt$ and taking the limit $dt\rightarrow 0$ gives the following leading order equation for $\sigma(t)$:
\begin{equation}
\label{sig}
\frac{d\sigma}{dt}=2 \overline{F}'(\nu)\sigma(t)  +2 \epsilon D(\nu).
\end{equation}

\begin{figure}[t!]
\begin{center} \includegraphics[width=9cm]{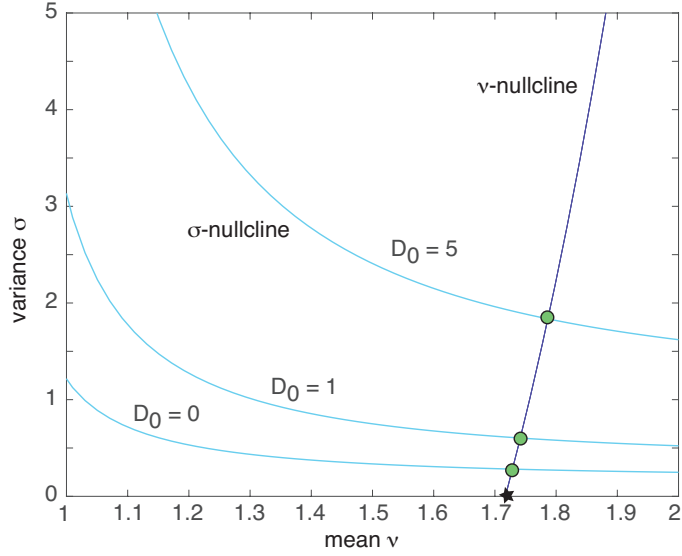} \end{center}
\caption{Plots of $\nu$-nullclines and $\sigma$-nullclines for equations (\ref{nudyn}) and (\ref{sigdyn}) with $f(\nu)=a\tanh(\gamma \nu)$. Various values of $D_0$ are considered with $\epsilon =0.1$, $\gamma=0.75$ and $a=2$.  Filled dots indicate the stable fixed point $(\nu^*,\sigma^*)$, which moves further away from  $(\nu_0,0)$ (indicated by the star) as $D_0$ increases from zero.}
\label{fig1}
\end{figure}

\begin{figure}[t!]
\begin{center} \includegraphics[width=9cm]{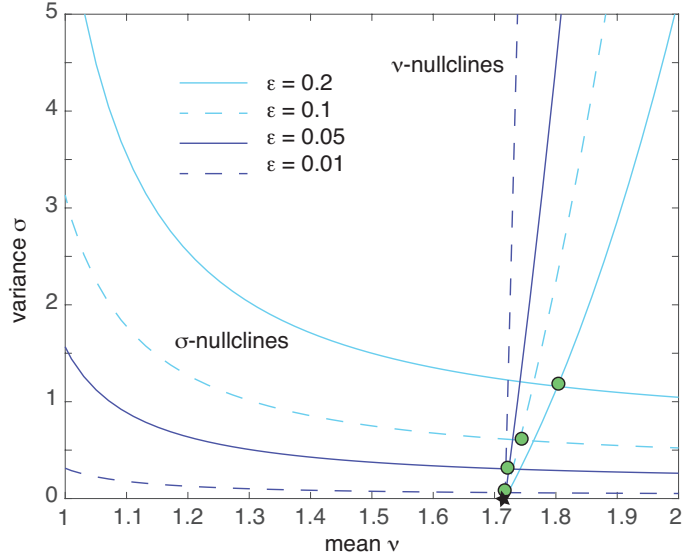} \end{center}
\caption{Plots of $\nu$-nullclines and $\sigma$-nullclines for equations (\ref{nudyn}) and (\ref{sigdyn}) with $f(\nu)=a\tanh(\gamma \nu)$. Various values of $\epsilon$ are considered with $D_0=1$, $
\gamma=0.75$ and $a=2$. Filled dots indicate the stable fixed point $(\nu^*,\sigma^*)$ with $(\nu^*,\sigma^*)\rightarrow (\nu_0,0)$ as $\epsilon \rightarrow 0$.}
\label{fig2}
\end{figure}

\noindent{\bf Example.}  Consider the stochastic hybrid system given by equations (\ref{nn1}) and (\ref{nn2}). Differentiating $\Lambda_0$ using equation (\ref{Rlam2}) and setting $p=0$, $x=\nu$ gives
\begin{equation}
\label{nudyn}
\frac{d\nu}{dt}=-\nu+f(\nu)+\frac{\epsilon}{2}\sigma (t)f''(\nu),
\end{equation}
which is the first-order correction to the rate equation (\ref{rate2}). In order to determine the diffusion function $D(\nu)$ in equation (\ref{sig}), we have to solve equation (\ref{Z}) for the given system:
\begin{eqnarray*}
\fl f(x)Z(n-1,x)+(n+1)Z(n+1,x) -(f(x)+n)Z(n,x)=(f(x)-1)\rho(n,x),
\end{eqnarray*}
with $ \rho(n,x)= {f(x)^n}\e^{-f(x)}/n!$.
It can be checked that this has the solution
\begin{equation}
Z(n,x)=(n-f(x))\rho(n,x),
\end{equation}
Hence, from equation (\ref{DD}) and properties of the Poisson distribution
\begin{equation}
\fl D(x)-D_0= \sum_n (n-x)(n-f(x))\rho(n,x)=\langle n^2\rangle -\langle n\rangle^2=f(x).
\end{equation}
Finally,
\begin{equation}
\label{sigdyn}
\frac{d\sigma}{dt}=-2 (1-f'(\nu))\sigma(t)  +2 \epsilon (f(\nu)+D_0).
\end{equation}
In Figs. \ref{fig1} and \ref{fig2} we illustrate how the $O(\epsilon)$ corrections affect the long-time dynamics of the mean and variance. For concreteness, we take $f(\nu)=a\tanh(\gamma \nu)$ with $a,\gamma$ chosen so that the $\epsilon =0$ case has an unstable fixed point at $(\nu,\sigma)=(0,0)$ and a stable fixed point at $(\nu,\sigma)=(\nu_0,0)$ with $\nu_0>0$. For $\epsilon >0$ we find that the fixed point shifts in the phase plane, that is, $(\nu_0,0)\rightarrow (\nu^*,\sigma^*)$ with $\nu^*>\nu_0$ and $\sigma^*>0$.

\subsection{Weak-coupling} So far we have discussed the weak-noise analog of the semiclassical limit in quantum path integrals. Another major perturbation scheme, common to both quantum and stochastic path integrals, is the weak-coupling limit. In order to formulate the basic idea, consider the following SDE \cite{Chow15}:
\begin{equation}
\label{OU+}
dX=-\gamma X+\epsilon f(X)+\sqrt{2\epsilon D(X)}dW,
\end{equation}
which involves a nonlinear perturbation of an Ornstein-Uhlenbeck process.
The corresponding path integral action is
\begin{equation}
S[x,p]=\int_0^t\left (p[\dot{x}+\gamma x]-\epsilon [pf(x)+p^2D(x)]\right )d\tau.
\end{equation}
Define the free Green's function or propagator $G_0(t,t')$ according to
\begin{equation}
\frac{dG_0}{dt}+\gamma G_0(t,t')=\delta(t-t').
\end{equation}
This has the solution
\begin{equation}
G_0(t,t')=H(t-t')\e^{-\gamma (t-t')},
\end{equation}
where $H(t)$ is the left continuous Heaviside step function: $H(0)=0$ for $t\leq 0$ and $H(1)=1$ for $t>0$. The choice $H(0)=0$ is consistent with the Ito condition for the SDE, and ensures that $X(t)$ is uncorrelated with $W(t)$. (Other forms of stochastic calculus would be obtained using different choices of $H(0)$. For example, $H(0)=1/2$ would  correspond to Stratonovich calculus.)

Given the definition of $G_0$, the action can be rewritten in the form
\begin{equation}
\fl S[x,p]=\int_0^t\int_0^td\tau\, d\tau'  p(\tau)G_0(\tau,\tau')^{-1} x(\tau') -\epsilon \int_0^t [pf(x)+p^2D(x)]d\tau,
\end{equation}
with the first term on the right-hand side identified as the so-called free action $S_0$.
Substituting into the associated generating functional gives
\begin{eqnarray}
\fl & Z[{J},\widetilde{J}] \sim \int  D[x]D[p]\exp \left (-S[x,p]+\int_0^t \left [x(\tau)\widetilde{J}(\tau)+J(\tau)p(\tau)\right ]d\tau\right )\\
\fl &\sim \int  D[x]D[p]\exp \left (-S_0[x,p]-\epsilon \Delta S[x,p]+\int_0^t \left [x(\tau)\widetilde{J}(\tau)+J(\tau)p(\tau)\right ]d\tau \right )\nonumber \\
\fl &\sim \e^{-\epsilon \Delta S[\delta/J(\tau),\delta/\delta \widetilde{J}(\tau)]}\int  D[x]D[p]\exp \left (-S_0[x,p]+\int_0^t \left [x(\tau)\widetilde{J}(\tau)+J(\tau)p(\tau)\right ]d\tau \right )\nonumber\\
\fl &\sim  \e^{-\epsilon \Delta S[\delta/J(\tau),\delta/\delta \widetilde{J}(\tau)]}\exp \left (\int_0^t\int_0^td\tau\, d\tau'  \widetilde{J}(\tau)G_0(\tau,\tau')^{-1} J(\tau') \right ).\nonumber
\end{eqnarray}
The final step follows from discretizing the path integral, noting that $p$ is an imaginary variable (due to the Wick rotation), and performing the complex Gaussian integral. This establishes the well-known result that for small $\epsilon$ (weak coupling), one can express moments of the full system as $\epsilon$ series expansions in products of free propagators (after setting $J=\widetilde{J}=0$).
Further details can be found in Ref. \cite{Chow15}.

A major challenge is identifying the analog of the weak-coupling limit for stochastic hybrid systems, where the action (\ref{actM0}) is defined in terms of the principal eigenvalue $\Lambda_0$. Motivated by the analysis of SDEs, suppose that the SDE (\ref{pd}) can be written in the form
\begin{equation}
dX=(-\gamma x+\epsilon f(n,x))dt+\sqrt{2\epsilon D(n,x)}dW,
\end{equation}
where the linear term is $n$-independent. Substituting into the eigenvalue equation (\ref{eigR}) gives
\begin{eqnarray}
& \sum_{m\geq 0}Q_{nm}(x) R(m,x,p)-(\gamma px +\Lambda(x,p))R(n,x,p)\nonumber \\
&\qquad =-\epsilon \left (p f(n,x)+p^2 D(n,x)\right )R(n,x,p).
\end{eqnarray}
Note that we are not necessarily considering a weak-noise limit, since the switching rates are not required to be $O(1/\epsilon)$. This suggests that we would need to sum over all $\mu$ in the path integral (\ref{G}). However, $\overline{R}_{\mu}(n,x,p)=\bar{r}_{\mu}(n,x) +O(\epsilon)$, where $\bar{r}_{\mu}$ is a left eigenvector of the matrix ${\bf A}$. Therefore, if we take $P(n_0,x,0)=\rho(n_0,x_0)$ with $\sum_{n_0}\bar{r}_{\mu}(n,x))\rho(n,x)=\delta_{\mu,0}$, then the contributions from non-principal components are $O(\epsilon)$. In addition, we will assume that the spectral gap is sufficiently large so that they decay rapidly compared to the principal component. This then allows us to restrict the sum over $\mu$ in the path integral (\ref{G}) to $\mu=0$. 

Introducing the perturbation expansions
\begin{equation}
R=R_0+\epsilon R_1+\epsilon^2 R_2+\ldots,\quad \Lambda=\Lambda_0+ \epsilon \Lambda_1+\epsilon^2 \Lambda_2+\ldots,
\end{equation}
plugging into the eigenvalue equation, and collecting terms at each power of $\epsilon$ generates a hierarchy of equations. In particular,
\numparts
\begin{eqnarray}
\label{hiera}
\fl & \sum_{m\geq 0}\L_{nm} R_0(m,x,p)=0,\\
\label{hierb}
\fl &\sum_{m\geq 0}\L_{nm}R_1(m,x,p)=[\Lambda_1(x,p) -p f(n,x)-p^2 D(n,x)] R_0(n,x,p),\\
\fl & \sum_{m\geq 0}\L_{nm}R_2(m,x,p)=[\Lambda_1(x,p) -p f(n,x)-p^2 D(n,x)]R_1(n,x,p)+\Lambda_2(x,p)R_0(x,p),\nonumber \\
\fl &
 \label{hierc}
\end{eqnarray}
\endnumparts
where
\begin{equation}
\L_{nm}=Q_{nm}(x)-(\Lambda_0(x,p)+\gamma px )\delta_{n,m}.
\end{equation}
In solving these equations we require that the eigenvector $R$ is positive so that we can identify $\Lambda$ with the principal eigenvalue. The first equation then implies
\begin{equation}
\Lambda_0=-\gamma p x,\quad R_0(m,x,p)=\rho(m,x),
\end{equation}
where $\rho$ is the positive right eigenvector of the (irreducible) matrix ${\bf Q}$. Next, summing both sides of equation (\ref{hierb}) with respect to $n$ using $\sum_nQ_{nm}=0$ and the explicit expressions for $\Lambda_0,R_0$ gives
\numparts
\begin{equation}
\Lambda_1(x,p)=p\overline{f}(x)+p^2\overline{D}(x),
\end{equation}
and $R_1$ is the solution to the equation
\begin{equation}
\fl  \sum_{m\geq 0}Q_{nm}(x)R_1(m,x,p)=\left (p[\overline{f}(x)-f(n,x)]+p^2[\overline{D}(x)-D(n,x)]\right )\rho(n,x).
 \end{equation}
\endnumparts
The latter has a solution from the Fredholm alternative theorem, i.e. both sides vanish when summed with respect to $n$. Moreover, the solution is unique if we impose the additional constraint $\sum_n R_1(n,x,p)=0$. Finally, summing both sides of equation (\ref{hierc}) with respect to $n$ yields
\begin{equation}
\Lambda_2(x,p)=\sum_{n\geq 0} [p f(n,x)+p^2 D(n,x)]R_1(n,x,p).
\end{equation}

We now substitute the perturbation expansion of the principal eigenvalue into the action (\ref{actM0}):
\begin{equation}
S[x,p]=\int_0^t\left \{p(\dot{x}+\gamma x)-\epsilon [p\overline{f}(x)+p^2\overline{D}(x)]+O(\epsilon^2) \right \}d\tau.
\end{equation}
Following along identical lines to the weak coupling expansion of SDEs, the generating functional for the stochastic hybrid system takes the form
\begin{eqnarray}
\fl & Z[{J},\widetilde{J}]  \sim  \e^{-\epsilon \Delta S[\delta/J(\tau),\delta/\delta \widetilde{J}(\tau)]}\exp \left (\int_0^t\int_0^td\tau\, d\tau'  \widetilde{J}(\tau)G_0(\tau,\tau')^{-1} J(\tau') \right ),\nonumber
\end{eqnarray}
with
\begin{equation}
\Delta S[x,p]=p\overline{f}(x)+p^2\overline{D}(x)+ \epsilon \Lambda_2(x,p)+O(\epsilon^2) .
\end{equation}
Note that the leading order contribution to $\Delta S$ is identical to $\Delta S$ for the SDE (\ref{OU+}) under the mappings $f(x)\rightarrow \overline{f}(x)$ and $D(x)\rightarrow \overline{D}(x)$. Hence, the term $\Lambda_2$ represents the leading order contribution from fluctuations beyond the Gaussian approximation. 

The above construction relies on the particular piecewise SDE given by equation (\ref{OU+}). One possible scenario where such an equation could arise is transcription of mRNA by an autoregulatory network in a slowly switching environment \cite{Hufton16,Bressloff17a}. In this case $x$ would be the mRNA concentration, $\gamma$ would represent the relatively fast degradation rate of mRNA, while $f_n(x)$ would be the nonlinear rate of synthesis when the environment is in state $n$. The source of the nonlinearity would be the fast binding/unbinding of transcriptional factor proteins synthesized by the mRNA.

\section{Conclusion}

In this paper we used operator methods borrowed from quantum mechanics to develop a path integral formulation of stochastic hybrid systems. These methods include bra-ket representations of Hilbert spaces, annihilation and creation operators, and position-momentum operators. The derivation showed how the eigenvalue equations (\ref{eigR}) and (\ref{eigS}) arise naturally in the construction of the hybrid path integral. An application of the Perron-Frobenius theorem then established how the principal eigenvalue $\Lambda_0$ plays the role of a Hamiltonian in determining least action paths. Unfortunately, there are few examples where explicit expressions for $\Lambda_0$ can be obtained unless the size of the Markov chain is sufficiently small so that brute force methods can be used. Therefore, we also explored various perturbation methods. First, we suggested one potential scheme for approximating $\Lambda_0$ in the case of large Markov chains, namely, the Ritz variational method. However, this requires a judicious choice of test functions. Second, we applied well-known approximations of quantum path integrals in the semiclassical limit to obtain corresponding approximations of the hybrid path integral and associated moment generating functional in the weak noise limit. Finally, we showed how the analog of a weak-coupling limit can be obtained by taking the deterministic contribution of the piecewise dynamics to be of the form $F(n,x)=-\gamma x+\epsilon f(n,x)$. Using analogies with weakly perturbed Ornstein-Uhlenbeck SDEs, we showed how the moment generating functional of a stochastic hybrid system can be used to develop a systematic perturbation expansion of moments in terms of products of free propagators. This suggests that there exists a systematic diagrammatic approach to analyzing moment equations for hybrid systems, which we hope to develop further elsewhere. It would also be interesting to explore generalizations of the Perron-Frobenius theorem that provide conditions for the existence of a spectral gap and a positive eigenvector in the case of semi-infinite Markov chains. Another future direction would be to consider stochastic hybrid PDEs as analogs of reaction-diffusion master equations.

\end{document}